\documentclass[referee, useAMS,usenatbib]{mn2e}
\usepackage{graphicx,graphics,amsmath,amssymb,mathrsfs}

\newcommand{\beq}{\begin{equation}}
\newcommand{\bea}{\begin{array}}
\newcommand{\beqa}{\begin{eqnarray}}
\newcommand{\eeq}{\end{equation}}
\newcommand{\eea}{\end{array}}
\newcommand{\eeqa}{\end{eqnarray}}

\newcommand{\bfr}{\mbox{\boldmath $r$}}
\newcommand{\bfv}{\mbox{\boldmath $v$}}
\newcommand{\bfT}{\mbox{\boldmath $T$}}
\newcommand{\bfQ}{\mbox{\boldmath $Q$}}

\newcommand{\bbxi}{\mbox{\boldmath $\xi$}}
\newcommand{\dmath}{\mbox{${\mathrm d}$}}

\newcommand{\p}{\mbox{$\partial$}}
\newcommand{\mbul}{\mbox{$M_\bullet$}}                                      
\newcommand{\im}{{\mathrm i}}
\newcommand{\e}{{\mathrm e}}

\begin{document}

\title[Counter--rotating stellar discs around a massive black hole]{Counter--rotating stellar discs around a massive black hole: self--consistent, time--dependent dynamics}

\author[Touma \& Sridhar]{J. R. Touma$^{1,3}$ and S. Sridhar$^{2,4}$   \\
  $^{1}$ Department of Physics, American University of Beirut, PO Box 11-0236, Riad El-Solh, Beirut 11097 2020, Lebanon \\  
  $^{2}$ Raman Research Institute, Sadashivanagar, Bangalore 560 080, India \\
  $^{3}$ jt00@aub.edu.lb   
  $^{4}$ ssridhar@rri.res.in \\
} 
\maketitle

\begin{abstract} We formulate the collisionless Boltzmann equation
(CBE) for dense star clusters that lie within the radius of influence
of a massive black hole in galactic nuclei. Our approach to these
nearly Keplerian systems follows that of Sridhar and Touma (1999):
Delaunay canonical variables are used to describe stellar orbits and
we average over the fast Keplerian orbital phases. The stellar
distribution function (DF) evolves on the longer time scale of
precessional motions, whose dynamics is governed by a Hamiltonian,
given by the orbit--averaged self--gravitational potential of the
cluster. We specialize to razor--thin, planar discs and consider two
counter--rotating (``$\pm$'') populations of stars. To describe discs
of small eccentricities, we expand the $\pm$ Hamiltonian to fourth
order in the eccentricities, with coefficients that depend
self--consistently on the $\pm$ DFs. We construct approximate $\pm$
dynamical invariants and use Jeans' theorem to construct
time--dependent $\pm$ DFs, which are completely described by their
centroid coordinates and shape matrices. When the centroid
eccentricities are larger than the dispersion in eccentricities, the
$\pm$ centroids obey a set of 4 autonomous equations ordinary
differential equations. We show that these can be cast as a
two--degree of freedom Hamiltonian system which is nonlinear, yet
integrable. We study the linear instability of initially circular
discs and derive a criterion for the counter--rotating instability. We
then explore the rich nonlinear dynamics of counter--rotating discs,
with focus on the variety of steadily precessing eccentric
configurations that are allowed. The stability and properties of these
configurations are studied as functions of parameters such as the disc
mass ratios and angular momentum.  \end{abstract}


\section{Introduction}

Galactic nuclei have massive black holes and dense clusters of stars,
whose structural and kinematic properties appear to be correlated with
global galaxy properties \citep{geb96, fm00, geb00}.
These correlations are probably the relics of the formation and evolution of
the galaxy and its central black hole \citep{ric98, hq11}. The
dynamics of star clusters around massive black holes involves physical
processes under extreme conditions, because of the high stellar
densities, large velocities and short time scales \citep{ale2005,
mer2006}. For a star cluster with 1--dimensional velocity dispersion
$\sigma$, and black hole of mass $\mbul$, the radius of influence of
the black hole is traditionally defined as $r_{\rm h} =
G\mbul/\sigma^2\,$. Within the radius of influence, i.e. for $r <
r_{\rm h}$, the dynamics of stars is dominated by the gravitational
attraction of the black hole. When general relativistic effects are
weak enough, the orbital dynamics is a perturbation of the Kepler
problem. \citet{st99} argued that the semi--major axis of stellar
orbits would be a secularly conserved quantity, and that this greater
integrability would facilitate the existence of asymmetric stellar
distributions. The secular dynamical evolution of nearly Keplerian
systems such as stellar clusters surrounding black holes in galactic
nuclei, cometary clouds or planetesimal discs were studied by
\citet{ttk09}.
 
For most galaxies, it is difficult to observe details of the stellar
distribution for $r < r_{\rm h}\,$.\footnote{$r_{\rm h} \sim 10\,{\rm
pc}$ for $\sigma \sim 200\,{\rm km}\,{\rm s}^{-1}\,$ and $\mbul \sim
10^8\, M_\odot $.} Therefore, observations of the nuclear regions of
our Galaxy and M31 assume special importance, because these are the
nearest large, normal galaxies for which it is possible to get a great
deal of photometric, kinematic and spectral information about the
stars. There is evidence for a black hole of mass $\sim 4\times 10^6\,
M_\odot$ at the Galactic center, with dense clusters of stars orbiting
it \citep{geg2010}. Among these, there is a population of about 200
young stars \citep{pau2006}, about half which probably belong to a
rotating disc which is highly warped \citep{lb2003, lu2006,
lu2009}. The remaining young stars appear to be members of a
counter--rotating population which is thicker than and inclined to the
warped disc \citep{gen2003, pau2006, bar2009, bar2010}. The nucleus of
M31 has a lopsided double--peaked distribution of stars orbiting a
black hole of mass $\sim 10^8\, M_\odot$, with the brighter peak
off--centered, and the fainter one centered close to the black hole
\citep{lds74, lau93, lau98, kb99}. \citet{tre95} proposed that the
off--centered peak marks the location in a stellar disc corresponding
to the aligned apoapsides of several eccentric stellar
orbits. Following this proposal, more detailed stellar dynamical
models of the eccentric disc have been proposed \citep{bac01, ss01,
ss02, pt03, ben05}. It is a very interesting fact that for both
galaxies, these extraordinary stellar dynamical structures have $r <
r_{\rm h}$.

An alternative and useful way of seeing the significance of the radius
of influence is this: a self--gravitating star cluster that lies
within $r_{\rm h}$ has mass $M < \mbul$. The orbit of a star may be
thought of as a slowly evolving Keplerian ellipse of fixed semi--major
axis (with the central mass at one focus), whose dynamics is governed
by a Hamiltonian which is the orbit--averaged gravitational potential
due to the star cluster \citep{st99}. This slow secular evolution time
scale is larger than the typical Keplerian orbital time by the large
factor $(\mbul/M)\,$. Slow modes of Keplerian discs were first
explored by \citet{sst99, tre01}. Orbital dynamics has been analyzed
and classified for the cases of non-axisymmetric planar \citep{st99},
and triaxial cluster potentials \citep{mv99, ss00, pm01, mv11}. One 
purpose of classifying stellar orbits is to be able to construct
stellar distribution functions (DFs) that can reproduce the photometry
and kinematics of stars around massive black holes in galactic
nuclei. When the time scales under consideration are smaller than the
relaxation times, the DF obeys the collisionless Boltzmann equation
(CBE); see \citet{bt08}. In this paper we begin with a formulation of
the CBE for the self--consistent, slow secular dynamics of star
clusters within the radius of influence of massive black holes.

The warped discs at the Galactic center are mutually
counter--rotating. The stellar dynamical model of \citet{ss02} for the
lopsided stellar disc in the nucleus of M31 included a few percent of
the stars on counter--rotating orbits. Here, it was proposed that the
lopsidedness of the nuclear disc of M31 could have been excited by the
counter--rotating instability, due to the accretion of a globular
cluster that spiraled in due to dynamical friction.  This proposal was
motivated by the work of \citet{tou02}, which suggested that even a
small fraction of mass in counter--rotating orbits could excite a
linear lopsided instability. \citet{ttk09} examine a secularly
unstable system of counterrotating discs, and follow the unfolding and
saturation of the instability into a global, uniformly precessing,
lopsided ($m=1$) mode. Counter--rotating streams of matter in a
self--gravitating disc are known to be unstable to lopsided modes
\citep{zh78, ara87, saw88, ms90, pp90, sm94, ljh97, ss10}. Mass
accretion in galactic nuclei will be such that the sense of rotation
of infalling material will be uncorrelated with the rotation of
pre--existing material surrounding the central black hole. In other
words, in the course of the evolution of a galaxy, having
counter--rotating systems in its nucleus is probably generic.

The main goal of this paper is to formulate and analyze the
time--dependent dynamics of counter--rotating stellar discs, which lie
within the radius of influence of the black hole. In \S~2 we discuss
the slow dynamics of nearly Keplerian star clusters by using the
Delaunay action--angle variables, and average over the fast orbital
phase. We present the CBE governing the collisionless evolution of the
stellar DF. The self--consistent Hamiltonian is the orbit--averaged
gravitational potential due to the star cluster, where softened
gravity is used. The CBE is then formulated for razor--thin, planar
discs, where the phase space is seen to be topologically equivalent to
a 2--sphere. In \S~3 we consider counter--rotating (``$\pm$'') discs
with fixed semi--major axes. Here, it proves convenient to write
separate CBEs for the $\pm$ DFs in the $\pm$ phase spaces. We consider
discs of small eccentricities, and introduce "cartesian-type"
canonical variables. The ring--ring interaction potential is expanded
to 4th order in the eccentricities for discs with the same semi--major
axes, using results from \citet{mt11} which are elaborated upon in
Appendix~A. Using this, the $\pm$ Hamiltonians are expressed in terms
of the $\pm$ DFs with coefficients that depend on both $\pm$ DFs. In
\S~4 we construct time--dependent DFs for the $\pm$ discs. The method
used is to first seek approximate dynamical invariants for the
time--dependent dynamics of the gravitationally coupled $\pm$ discs,
and then use Jeans' theorem to construct time--dependent DFs. The
isocontours of the DFs are ellipses centered on moving origins in the
$\pm$ phase spaces. The coordinates of the centers (referred to as
``centroids'') contain information about the mean eccentricities and
periapse orientations of the $\pm$ DFs. Information about the
dispersions of eccentricities and periapse orientations is contained
in the positive--definite $2\times 2$ ``shape matrices'' which
describe the elliptical isocontours. When the centroid eccentricities
are larger than the dispersion in eccentricities, the centroid
dynamics is independent of the shape dynamics; however, the shape
dynamics is driven by the centroid dynamics. In \S~5 we show that the
coupled dynamics of the $\pm$ centroids is a two--degree of freedom
Hamiltonian system which is nonlinear, yet integrable.  We then study
the linear instability of initially circular discs and derive a
criterion for the instability. The rest of the section is an
exploration of the rich nonlinear dynamics of counter--rotating discs,
with focus on the variety of steadily precessing eccentric
configurations that are allowed. The stability and properties of these
configurations are studied as functions of parameters such as the disc
mass ratios and angular momentum. Summary and conclusions are offered
in \S~6.
 
\section{Collisionless evolution of nearly Keplerian stellar systems}

We consider a stellar system around a central object of mass
$\mbul\,$. Over times shorter than the relaxation times, the system is
effectively collisionless. Then the stellar system may be thought of
as composed of an infinite number of stars, each of infinitesimal
mass, with total mass in stars equal to $M$. Stellar orbits are
governed by the Newtonian gravity of the central mass, as well as the
mean--field gravitational potential of all the stars. When $(M/\mbul)
\ll 1$, it is useful to regard the dynamics as a perturbation of the
Kepler problem. Thus the orbit of a star may be thought of as a slowly
evolving Keplerian ellipse of fixed semi--major axis, with the central
mass at one focus. This slow secular evolution time scale is larger
than the typical Keplerian orbital time by the large factor
$(\mbul/M)\,$.

Each star is a moving point--like object in the $6$--dimensional phase
space, $\left(\bfr, \bfv\right)$, where $\bfr$ is its position with
respect to the central mass, and $\bfv$ is its velocity. Since the
dynamics of a star is nearly Keplerian, it is useful to employ the
Delaunay variables, which are action--angle variables for the Kepler
problem. The Delaunay variables, $\{I, L, L_{z}\,;\, w,
g, h\}$, are a set of action and angle variables for the Kepler
problem \citep{st99,bt08}. The three actions are: $I \,=\,\sqrt{G\mbul
a}\,$ where $a$ is the semi--major axis; $L$, which is the magnitude
of the orbital angular momentum; and $L_{z}$, which is the
$z$--component of the orbital angular momentum. The angles conjugate
to them are, respectively: $w$, the orbital phase; $g$, the angle to
periapse from the ascending node; and $h$, the longitude of the
ascending node.

In the absence of self--gravity, the motion of the star is purely
Keplerian: the orbital phase $w$ advances steadily at a rate equal to
the Keplerian orbital frequency, whereas the other five Delaunay
variables are constant in time. However, self--gravity contributes to
a slow variation of the Delaunay variables. Let $\widetilde{H}$ be the
total gravitational potential seen by a star, averaged over the
Keplerian orbital phase of the concerned star. Then $\widetilde{H}
\,=\, \widetilde{H}(I, L, L_z, g, h, t)$, where we have allowed for a
slow time dependence. Since $\widetilde{H}$ is --- by definition ---
independent of $w$, the conjugate momentum, $I$, is a conserved
quantity; the star's orbit can be imagined to be a slowly deforming ``Gaussian
ring'' of fixed semi--major axis, with the central mass stationary at
one focus.\footnote{This is a restatement of the well--known result in
planetary dynamics that the semi--major axis, $a$, is a secular
invariant. Henceforth we use $a$ instead of $I$.} The slow secular
evolution of the other Delaunay variables is given by

\beq
\frac{\dmath L}{\dmath t} \;=\; -\,\frac{\p \widetilde{H}}{\p g}\,,\qquad  
\frac{\dmath g}{\dmath t} \;=\; \frac{\p \widetilde{H}}{\p L}\,;\qquad
\frac{\dmath L_z}{\dmath t} \;=\; -\,\frac{\p \widetilde{H}}{\p h}\,,\qquad  
\frac{\dmath h}{\dmath t} \;=\; \frac{\p \widetilde{H}}{\p L_z}\,.
\label{deleom}
\eeq 
 
Let the stellar system be described by a distribution function (DF),
$f(a, L, L_z, g, h, t)$, where $f\,\dmath a\,\dmath L\,\dmath
L_z\,\dmath g\,\dmath h$ is the mass in the element $\left(\dmath
a\,\dmath L\,\dmath L_z\,\dmath g\,\dmath h\right)$. The collisionless
Boltzmann equation (CBE) which describes the time evolution of the DF
is

\beq
\frac{\dmath f}{\dmath t} \;\equiv\; \frac{\partial f}{\partial t} + \left[f, \widetilde{H} \right] \;=\; 0\,,
\label{cbefull}
\eeq 

\noindent
where
 
\beq\left[f, \widetilde{H} \right] \;=\; \frac{\partial f}{\partial g}\,\frac{\partial \widetilde{H}}{\partial L} \;-\; \frac{\partial f}{\partial L}\,\frac{\partial \widetilde{H}}{\partial g} \;+\; \frac{\partial f}{\partial h}\,\frac{\partial \widetilde{H}}{\partial L_z} \;-\; \frac{\partial f}{\partial L_z}\,\frac{\partial \widetilde{H}}{\partial h}\,,
\label{pbfull}
\eeq

\noindent
is the Poisson Bracket between $f$ and $\widetilde{H}$, defined in the $(L, L_z, g, h)$ phase space. The Hamiltonian is 

\beqa
\widetilde{H} \;=\; \Phi_{\rm ext}(a, L, L_z, g, h, t) &-& G \int \,\dmath a'\,\dmath L'\,\dmath L'_z\,\dmath g'\,\dmath h'\;f(a', L', L'_z, g', h', t)\;\times\nonumber\\[1em]
&&\qquad\qquad\times\;\Psi(a, L, L_z, g, h; a',L',L'_z, g', h'),
\label{cbehamfull}
\eeqa 

\noindent
where $\Phi_{\rm ext}$ is the gravitational potential due to an external
source, averaged over the star's Keplerian orbital phase. The second term
is the self--gravitational potential of the star cluster; the quantity,
 
\beq
\Psi(a, L, L_z, g, h; a',L',L'_z, g', h') \;=\; \oint\oint\,\frac{dw}{2\pi}\,\frac{dw'}{2\pi}\,\frac{1}{\left[\,\left|{\bf r} \,-\, {\bf r'}\right|^2 \;+\; b^2\;\right]^{1/2}}\;,
\label{phifull}
\eeq 

\noindent is proportional to the mutual gravitational potential energy
between two particles, each of unit mass, averaged over their
Keplerian orbital phases. Note that the gravitational interaction
between the stars has a softening length $b$, whereas that between the
central mass and a star is through an unsoftened Keplerian
potential. ``Softened gravity'' was introduced by \citet{mil71}, and
used as a surrogate for velocity dispersion in a stellar disc; more
recent work using softened gravity in the context of nearly Keplerian
systems are \citet{tou02, ttk09}. Since each star is usefully imagined
to be a ``Gaussian ring'' in secular dynamics, we refer to $\Psi$ as
the ring--ring interaction function. At each value of the semi--major
axis, $a$, equations~(\ref{cbefull})---(\ref{cbehamfull}) provide a
self--consistent description of slow secular dynamics in the $(L, L_z,
g, h)$ phase space.
 
\subsection{The CBE for razor--thin discs}

When motion is restricted to a plane, the description of secular
dynamics simplifies by reduction of the phase space by two
dimensions. Let this plane be chosen perpendicular to the
$z$--axis. The angles, $g$ and $h$, no longer have clear independent
meanings; rather it is the sum $(g + h)$ that is well--defined, and we
will henceforth refer to this as the angle $g$. Its conjugate variable is the scaled action
variable, $\,\ell = L_z/\sqrt{G \mbul a}\,$, which is equal to the
$z$--component of the angular momentum divided by the angular momentum
of a circular orbit of radius $a\,$: this definition makes $\ell$ a
normalized quantity: $-1\leq\ell\leq 1\,$. Then

\beq
\frac{\dmath g}{\dmath t} \;=\; \frac{\partial H}{\partial\ell}\,,\qquad\quad
\frac{\dmath \ell}{\dmath t} \;=\; -\frac{\partial H}{\partial g}\,.
\label{lgeom}
\eeq

\noindent
where the Hamiltonian $H(a,\ell, g, t) \;=\; \widetilde{H}/\sqrt{G \mbul a}\,$.
The DF, $f(a, \ell, g, t)$, satisfies the CBE:

\beq
\frac{\dmath f}{\dmath t} \;\equiv\; \frac{\partial f}{\partial t} + \left[f, H \right] \;=\; 0\,,
\label{cbe}
\eeq 

\noindent
where 
\beq\left[f, H \right] \;=\; \frac{\partial f}{\partial g}\,\frac{\partial H}{\partial \ell} \;-\; \frac{\partial f}{\partial \ell}\,\frac{\partial H}{\partial g}\,
\label{pborg}
\eeq

\noindent
is the Poisson Bracket between $f$ and $H$, defined in the $(\ell,g)$ phase space. $H$ is determined self--consistently through

\beq
H \;=\; \Phi_{\rm ext}(a, \ell, g, t) \;-\; \frac{G}{\sqrt{G \mbul a}} \int \dmath a'\dmath \ell' \dmath g' \,f(a',\ell',g',t)\,\Psi(a, \ell, g; a', \ell', g').
\label{cbeham}
\eeq 
Note that the kinetic energy and the Kepler part of the potential energy have been omitted because their sum which is equal to $-G \mbul/2 a$ is a constant. The $(\ell, g)$ phase space is, topologically speaking, a 2--dimensional sphere, with $\ell$ equal to the cosine of the colatitude and $g$ equal to the azimuthal angle: $\ell = \pm 1$ are located at the north and south poles respectively, whereas $\ell = 0$ corresponds to the equator. For each value of the semi--major axis, $a$, equations~(\ref{cbe})---(\ref{cbeham}) provide a self--consistent description of slow secular dynamics on this 2--sphere.

\section{Counter--rotating discs: formalism for small eccentricities}

In our study of counter--rotating discs we restrict attention to isolated ($\Phi_{\rm ext} = 0$) planar, razor--thin discs around a central mass. We also assume that all stars in a disc have the same semi--major axis: the $+$ disc has stars with semi--major axis equal to $a_{+}$ and $-$ disc has stars with semi--major axis equal to $a_{-}$. Thus we choose the DF to be of the form,

\beq
f(a, \ell, g, t) \;=\; \delta(a-a_{+})\,f^{+}(\ell, g, t) \;\;+\;\; \delta(a-a_{-})\,f^{-}(\ell, g, t),
\label{dftwopop}
\eeq

\noindent
where the $\delta$--functions fix the semi--major axes of the two populations, which are now described by two different DFs, $f^{+}$ and  $f^{-}$. These DFs obey separate CBEs:

\beq
\frac{\partial f^{+}}{\partial t} \;+\; \left[f^{+}, H^{+} \right] \;=\; 0\,; \qquad\quad
\frac{\partial f^{-}}{\partial t} \;+\; \left[f^{-}, H^{-} \right] \;=\; 0\,,
\label{cbepm}
\eeq

\noindent
where $H^{\pm}$ are the Hamiltonians acting on the $\pm$ populations, respectively. Each of $H^{+}$ and $H^{-}$ depends on both $f^{+}$ and $f^{-}$, leading to coupled dynamics of the $\pm$ populations. In fact,
 
\beqa
H^{+}(\ell, g, t) &=& -\,\frac{G}{\sqrt{G \mbul a_{+}}} \int \dmath \ell' \dmath g'\, f^{+}(\ell',g',t)\,\Psi(a_{+}, \ell, g; a_{+}, \ell', g') \;\;-\; \nonumber\\[2em]
&& -\,\frac{G}{\sqrt{G \mbul a_{+}}} \int \dmath \ell' \dmath g'\, f^{-}(\ell',g',t)\,\Psi(a_{+}, \ell, g; a_{-}, \ell', g'),
\label{hamplus} 
\eeqa

\noindent
and
 
\beqa
H^{-}(\ell, g, t) &=& -\,\frac{G}{\sqrt{G \mbul a_{-}}} \int \dmath \ell' \dmath g'\, f^{-}(\ell',g',t)\,\Psi(a_{-}, \ell, g; a_{-}, \ell', g') \;\;-\; \nonumber\\[2em] 
&& -\,\frac{G}{\sqrt{G \mbul a_{-}}} \int \dmath \ell' \dmath g'\, f^{+}(\ell',g', t)\,\Psi(a_{-}, \ell, g; a_{+}, \ell', g')\,.
\label{hamminus}
\eeqa

We now specialize to counter--rotating discs of small eccentricities:
let $f^{+}$ be the DF for a prograde population, which is concentrated
around $\ell=+1$, and $f^{-}$ be the DF for a retrograde DF which is
concentrated around $\ell=-1$. We are interested in recovering a
truncated model for the collective dynamics of these coupled
populations. We recall that the $(\ell, g)$ phase space is a
2--sphere, with $\ell$ equal to the cosine of the colatitude and $g$
equal to the azimuthal angle. The prograde and retrograde populations
we consider are concentrated at the north and south poles,
respectively. So it is convenient to use two different coordinate
patches to describe the two populations.\footnote{Our analysis is
limited to scenarios in which the two populations preserve their
identity as prograde and retrograde stars. In other words, the sign of
the orbital angular momentum of each star does not change.} Thus we
choose separate prograde and retrograde canonical variables
\beqa
I_{+} &=& 1 - \ell\,,\qquad\quad \theta_{+} \;=\; -g\;;\nonumber\\[1em]
I_{-} &=& 1 + \ell\,,\qquad\quad \theta_{-} = g\,.
\label{ithetadef}
\eeqa
\noindent We will also find it convenient to use the ``cartesian counterparts'' of the $(I, \theta)$ variables.
These are defined by
  
\beqa
x_{+} &=& \sqrt{2 I_{+}}\,\sin\theta_{+} \;=\; -\sqrt{2(1-\ell)}\,\sin g\,,
\nonumber\\[1em]
y_{+} &=& \sqrt{2 I_{+}}\,\cos\theta_{+} \;=\; \sqrt{2(1-\ell)}\,\cos g\,;
\label{xyplus}\\[2em]
x_{-} &=& \sqrt{2 I_{-}}\,\sin\theta_{-} \;=\; \sqrt{2(1+\ell)}\,\sin g\,, 
\nonumber\\[1em]
y_{-} &=& \sqrt{2 I_{-}}\,\cos\theta_{-} \;=\; \sqrt{2(1+\ell)}\,\cos g.
\label{xyminus}
\eeqa
\noindent
Here $x_{\pm}$ are new coordinates, and $y_{\pm}$ are new momenta for the $\pm$ populations. The
transformations from old to new variable are of course canonical and can be simply recovered with the help of the generating function $S(x_{\pm}, {\theta}_{\pm}) = (x_{\pm}^2/2) \cot {\theta}_{\pm}$. 

Before we plunge into a series of approximations that will yield the reduced dynamics, we further simplify the model disc, by further specializing to $\pm$ populations with the same semi-major axes: 

\beq
a_{+} \;=\; a_{-} \;=\; a_0\,.
\label{asame}
\eeq

\noindent The principal advantage of this restriction is that the
ring--ring interaction function, $\Psi$, can be described by fewer
constants, allowing us to develop the theory with less
clutter. However it may miss describing new phenomena when $a_{+} \neq
a_{-}$. We reiterate:

\begin{itemize}

\item We will study the planar secular dynamics of two counter--rotating stellar discs of small eccentricities, around a central massive object. 

\item All the stars are assumed to have the same (conserved) semi--major axis. 

\item Each star has a single degree of a freedom, namely the freedom to adjust its periapse orientation and conjugate eccentricity in response to collective (smooth) gravitational potential of all the other stars.

\item A star can, of course, change the sign of its orbital angular momentum and switch membership from the $+$ population to the $-$ population (or vice versa), while preserving its semi--major axis. However, this  freedom is not allowed in what follows and shall be relaxed in future considerations of this problem.

\end{itemize}

\subsection{Expansion of the ring--ring interaction function}

To work out the Hamiltonians $H^{+}$ and $H^{-}$, we expand the
ring--ring interaction function, $\Psi$, to $4^{\rm th}$ order in the
eccentricities of the rings\footnote{Here a ring is thought of as a
single Keplerian orbit which has been averaged over its fast orbital
phase. In the context of stellar dynamics, we can also imagine a ring
as a single Kepler orbit populated by many stars with the same
Keplerian orbital elements, except for their orbital phases which are
equally distributed over 2$\pi$.}. This expansion was developed by
\citet{mt11}, and is given in the Appendix~A. Let us define ${\bf e}$
and ${\bf e'}$, the eccentricity vectors characterizing two rings:

\beq 
{\bf e} \;=\; (e\cos g\,, \,e\sin g\,)\;;\qquad\quad {\bf e'} \;=\; (e'\cos g'\,, \,e'\sin g'\,)\,,
\label{eeprime}
\eeq
\noindent
with $e = \sqrt{1-{\ell}^2}$ and $e' = \sqrt{1-{\ell'}^{2}}$. In the expansion of $\Psi$ we drop terms of the following type: (i) terms that are independent of ${\bf e}$ because these do not contribute to the dynamics of the concerned ring; (ii) terms higher that $4^{\rm th}$ order in ${\bf e}, {\bf e}'$, because this is the accuracy we aim for. Then,

\beq
\sqrt{\frac{G}{\mbul a_{0}}}\,\Psi \;\;=\;\; \alpha e^2 \;+\; \beta {\bf e}\cdot{\bf e'} \;+\; \gamma e^2 e'^2 \;+\; \lambda\left({\bf e}\cdot{\bf e'}\right)^{2} \;+\; \kappa\left({\bf e}\cdot{\bf e'}\right)e'^2 \;+\; \chi e^4 \;+\; \kappa e^2\left({\bf e}\cdot{\bf e'}\right)\,,
\label{phiexp}
\eeq

\noindent where the coefficients $(\alpha, \beta, \gamma, \lambda,
\kappa, \chi)$ are functions of $a_0$ and $b$, and are given in terms
of the softened Laplace coefficients (see the Appendix~A). Note that
$3^{\rm rd}$ order terms are absent. Each of the ${\bf e}$ and ${\bf
e'}$ can belong to either the $+$ or $-$ population, so there are four
possibilities. We first express ${\bf e}$ in terms of $\left(x_{\pm},
y_{\pm}\right)$ accurate to $4^{\rm th}$ order. For the $+$ population
equations~(\ref{ithetadef}) and (\ref{xyplus}) give:

\beqa
e^2 &=& 1-\ell^2  \bea{cccc} =& 1 \;-\; \left(1-I_{+}\right)^2 &=&  x_{+}^2 \;+\; y_{+}^2 \;-\; \frac{1}{4}\left(x_{+}^2+ y_{+}^2\right)^2\eea \nonumber\\[1em]
e\cos g &=& \sqrt{2 I_{+}\left(1 - \frac{I_{+}}{2}\right)}\,\cos\theta_{+} \bea{cc}  =& y_{+} \sqrt{1 - \frac{1}{4}\left(x_{+}^2+ y_{+}^2\right)} \eea \nonumber\\[1em]
e\sin g &=& -\,\sqrt{2 I_{+}\left(1 - \frac{I_{+}}{2}\right)}\,\sin\theta_{+} \bea{cc} =& -\,x_{+}\sqrt{1 - \frac{1}{4}\left(x_{+}^2+ y_{+}^2\right)}\;.\eea
\label{exyplus}
\eeqa

\noindent
Similarly, for the $-$ population, equations~(\ref{ithetadef}) and (\ref{xyminus}) give:
\beqa
e\cos g &=& \sqrt{2 I_{-}\left(1 - \frac{I_{-}}{2}\right)}\,\cos\theta_{-} \bea{cc} =& y_{-} \sqrt{1 - \frac{1}{4}\left(x_{-}^2+ y_{-}^{2}\right)} \eea \nonumber\\[1em]
e\sin g &=& \,\sqrt{2 I_{-}\left(1 - \frac{I_{-}}{2}\right)}\,\sin\theta_{-} \bea{cc} =& \,x_{-}\sqrt{1 - \frac{1}{4}\left(x_{-}^2+ y_{-}^{2}\right)}\;. \eea
\label{exyminus}
\eeqa

\noindent
To express ${\bf e'}$ in terms of $\left(x'_{\pm}, y'_{\pm}\right)$ we  simply add the primes to the above expressions. Our next task is to obtain
expressions for $\Psi(+, +')$, $\Psi(+, -')$, $\Psi(-, +')$, and $\Psi(-, -')$ to the same order.\footnote{We employ an obvious shorthand: for instance, 
$\Psi(+, -')$ is the interaction function between two rings, one with 
parameters $\left(x_{+}, y_{+}\right)$ and the other with parameters
$\left(x'_{-}, y'_{-}\right)\,$.} We begin with $\Psi(+, +')$, by working out
the terms involved in the $(+,+')$ interactions. Using equation~(\ref{exyplus}), we have:

\beqa
e^2 &=& x_{+}^2 \;+\; y_{+}^2 \;-\; \frac{1}{4}\left(x_{+}^2 \,+\, y_{+}^2\right)^2\,,\nonumber\\[1em]
{\bf e}\cdot{\bf e'} &=& \left(x_{+}x'_{+} \,+\, y_{+}{y'}_{+}\right)\,\sqrt{\left(1 \,-\, \frac{x_{+}^2 \,+\, y_{+}^2}{4}\right)\left(1-\frac{{x'}_{+}^2 \,+\, {y'}_{+}^2}{4}\right)}\nonumber\\[1em]
&=& \left(x_{+} x'_{+} \,+\, y_{+}{y'}_{+}\right)\left[\,1 \;-\; \frac{x_{+}^2 \,+\, y_{+}^2 \,+\, {x'}_{+}^2 \,+\, {y'}_{+}^2}{8}\right] \;+\; \ldots\,.
\nonumber\\[1em]
e^2 {e'}^2 &=& \left(x_{+}^2 \,+\, y_{+}^2\right)\left({x'}_{+}^2 \,+\, {y'}_{+}^2\right) \;+\; \ldots\nonumber\\[1em]
\left({\bf e}\cdot {\bf e'}\right)^2 &=& \left(x_{+} x'_{+} +  y_{+}{y'}_{+}\right)^2 \;+\; \ldots\nonumber\\[1em]
e'^{2}\,{\bf e}\cdot{\bf e'} &=& \left({x'}_{+}^2+ {y'}_{+}^2\right)\left(x_{+} x'_{+} + y_{+}{y'}_{+}\right) \;+\; \ldots\nonumber\\[1em]
e^{4} &=& \left(x_{+}^2+ y_{+}^2\right)^{2} \;+\; \ldots\nonumber\\[1em]
e^{2}\,{\bf e}\cdot {\bf e'} &=& \left({x}_{+}^2+ {y}_{+}^2\right)\left(x_{+} x'_{+} + y_{+}{y'}_{+}\right) \;+\;\ldots\; .
\label{ppfirst} 
\eeqa

\noindent
Then, substituting equations~(\ref{ppfirst}) in (\ref{phiexp}), the interaction function between two prograde rings reduces to:

\beqa
\sqrt{\frac{G}{\mbul a_{0}}}\,{\Psi}(+, +') &=& \left[\alpha + \gamma \left({x'}_{+}^2 + {y'}_{+}^2\right) +  \lambda {x'}_{+}^2 \right]\,x_{+}^2 \nonumber\\[1em]
&+& \left[\alpha + \gamma \left({x'}_{+}^2 + {y'}_{+}^2\right) +  \lambda {y'}_{+}^2 \right]\,y_{+}^2 \;+\;  [2\lambda x'_{+}{y'}_{+}]\,x_{+}y_{+} \nonumber\\[1em]
&+& \left[ \beta {x'}_{+} + \left(\kappa -\frac{\beta}{8}\right){x'}_{+}\left({{x'}}_{+}^2 + {y'}_{+}^2\right)\right]\,x_{+} \nonumber\\[1em]
&+& \left[ \beta {y'}_{+} + \left(\kappa -\frac{\beta}{8}\right){y'}_{+}\left({x'}_{+}^2+  {y'}_{+}^2\right) \right]\,y_{+} \nonumber\\[1em]
&+&  \left[\left(\kappa -\frac{\beta}{8}\right) {x'}_{+}\right]\,x_{+}\left(x_{+}^2+ y_{+}^2\right) \nonumber\\[1em]
&+& \left[\left(\kappa -\frac{\beta}{8}\right) {y'}_{+}\right]\,y_{+}\left(x_{+}^2+ y_{+}^2\right) \;+\; \left(\chi -\frac{\alpha}{4}\right)\,\left(x_{+}^2+ y_{+}^2\right)^{2}\;,
\label{phippexp}
\eeqa

\noindent
where we have lumped all the dependences on primed quantities within the square brackets. Similarly, we compute quantities for the $(+, -')$ interactions using equations~(\ref{exyminus}). Then,
 
\beqa
\sqrt{\frac{G}{\mbul a_{0}}}\,{\Psi}(+, -') &=& \left[\alpha + \gamma \left({x'}_{-}^2 + {y'}_{-}^2\right) +  \lambda {x'}_{-}^2 \right]\,x_{+}^2 \nonumber\\[1em]
&+& \left[\alpha + \gamma \left({x'}_{-}^2 + {y'}_{-}^2\right) +  \lambda {y'}_{-}^2 \right]\,y_{+}^2 \;-\;  [2\lambda x'_{-}{y'}_{-}]\,x_{+}y_{+}  \nonumber\\[1em]
&-& \left[ \beta {x'}_{-} + \left(\kappa -\frac{\beta}{8}\right){x'}_{-}\left({{x'}}_{-}^2 + {y'}_{-}^2\right)\right]\,x_{+} \nonumber\\[1em]
&+& \left[ \beta {y'}_{-} + \left(\kappa -\frac{\beta}{8}\right){y'}_{-}\left({x'}_{-}^2 +  {y'}_{-}^2\right) \right]\,y_{+} \nonumber\\[1em]
&-&  \left[\left(\kappa -\frac{\beta}{8}\right) {x'}_{-}\right]\,x_{+}\left(x_{+}^2+ y_{+}^2\right) \nonumber\\[1em]
&+& \left[\left(\kappa -\frac{\beta}{8}\right) {y'}_{-}\right]\,y_{+}\left(x_{+}^2+ y_{+}^2\right) \;+\; \left(\chi -\frac{\alpha}{4}\right)\,\left(x_{+}^2+ y_{+}^2\right)^{2}\;.
\label{phipmexp}
\eeqa

\subsection{Self--consistency: Hamiltonians in terms of the DFs}

We can now compute $H^{+}$ by using equations~(\ref{phippexp}) and (\ref{phipmexp}) in (\ref{hamplus}). Then, obtaining an expression for $H^{-}$ is just a matter of switching signs: replace all the $+$ variables by the $-$ variables and vice versa. Putting together all these expansions, one finally recovers the Hamiltonians governing the prograde and retrograde populations:

\beqa
H^{\pm} &=& \frac{1}{2} A_{\pm} x^{2}_{\pm} \;+\; B_{\pm} x_{\pm} y_{\pm} \;+\; \frac{1}{2} C_{\pm} y^{2}_{\pm} \;+\; D_{\pm} x_{\pm} \;+\; E_{\pm} y_{\pm} \nonumber\\[1em]
&+& F_{\pm} x_{\pm}\left(x^{2}_{\pm} \,+\, y^{2}_{\pm}\right) \;+\; G_{\pm} y_{\pm}\left(x^{2}_{\pm} \,+\, y^{2}_{\pm}\right) + K \left(x^{2}_{\pm} \,+\, y^{2}_{\pm}\right)^{2}\,,
\label{hampm}
\eeqa

\noindent
where the coefficients are determined self--consistently in terms of the DFs $f^{+}$ and $f^{-}$ by,

\beqa
A_{\pm}(t) &=& -\,2 \int \dmath x_{\pm}\dmath y_{\pm}\,f^{\pm}(x_{\pm}, y_{\pm}, t)\left[\alpha + (\gamma+\lambda) x_{\pm}^2 + \gamma y_{\pm}^2\right]\nonumber\\[1em] && -\,2 \int \dmath x_{\mp} \dmath y_{\mp}\, f^{\mp}(x_{\mp}, y_{\mp}, t)\left[\alpha + (\gamma+\lambda)x_{\mp}^2 + \gamma y_{\mp}^2\right]\nonumber\\[2em]
B_{\pm}(t) &=& -\,2 \lambda \int \dmath x_{\pm} \dmath y_{\pm}\, f^{\pm}(x_{\pm}, y_{\pm}, t) x_{\pm} y_{\pm} \;+\; 2 \lambda \int \dmath x_{\mp} \dmath y_{\mp} f^{\mp}(x_{\mp},
y_{\mp}, t) x_{\mp} y_{\mp} \nonumber \\[2em]
C_{\pm}(t) &=& -\,2 \int \dmath x_{\pm} \dmath y_{\pm}\, f^{\pm}(x_{\pm}, y_{\pm}, t)  \left[\alpha + \gamma x_{\pm}^2 + (\gamma+\lambda) y_{\pm}^2 \right]\nonumber \\[1em] && -\,2 \int \dmath x_{\mp} \dmath y_{\mp}\, f^{\mp}(x_{\mp}, y_{\mp}, t)  \left[\alpha + \gamma x_{\mp}^2 + (\gamma+\lambda) y_{\mp}^2 \right] \nonumber \\[2em]
D_{\pm}(t) &=& -\,\int \dmath x_{\pm} \dmath y_{\pm}\, f^{\pm}(x_{\pm}, y_{\pm}, t) \left[\beta x_{\pm} + \left(\kappa -\frac{\beta}{8}\right) x_{\pm} \left(x_{\pm}^2 +  y_{\pm}^2\right) \right] \nonumber \\[1em]
&& +\, \int \dmath x_{\mp} \dmath y_{\mp}\, f^{\mp}(x_{\mp}, y_{\mp}, t) \left[\beta x_{\mp} + \left(\kappa -\frac{\beta}{8}\right) x_{\mp} \left(x_{\mp}^2 +  y_{\mp}^2\right) \right] \nonumber \\[2em]
E_{\pm}(t) &=& -\,\int \dmath x_{\pm} \dmath y_{\pm} \,f^{\pm}(x_{\pm}, y_{\pm}, t) \left[\beta y_{\pm} + \left(\kappa -\frac{\beta}{8}\right) y_{\pm} \left(x_{\pm}^2 +  y_{\pm}^2\right) \right] \nonumber \\[1em]
&& -\, \int \dmath x_{\mp} \dmath y_{\mp}\,f^{\mp}(x_{\mp}, y_{\mp}, t) \left[\beta y_{\mp} + \left(\kappa -\frac{\beta}{8}\right) y_{\mp} \left(x_{\mp}^2 +  y_{\mp}^2\right) \right] \nonumber \\[2em]
F_{\pm}(t) &=& - \left(\kappa -\frac{\beta}{8}\right) \int \dmath x_{\pm} \dmath y_{\pm}\,f^{\pm}(x_{\pm}, y_{\pm}, t) x_{\pm} \;+\; \left(\kappa -\frac{\beta}{8}\right) \int \dmath x_{\mp} \dmath y_{\mp}\, f^{\mp}(x_{\mp}, y_{\mp}, t) x_{\mp}\nonumber\\[2em]
G_{\pm}(t) &=& - \left(\kappa -\frac{\beta}{8}\right) \int \dmath x_{\pm} \dmath y_{\pm}\,f^{\pm}(x_{\pm}, y_{\pm}, t) y_{\pm} \;-\; \left(\kappa -\frac{\beta}{8}\right) \int \dmath x_{\mp} \dmath y_{\mp}\, f^{\mp}(x_{\mp}, y_{\mp}, t) y_{\mp}\nonumber\\[2em]
K  &=& - \left(\chi -\frac{\alpha}{4}\right) \int \dmath x_{+} \dmath y_{+}\,f^{+} \;-\; \left(\chi -\frac{\alpha}{4}\right) \int \dmath x_{-} \dmath y_{-}\,f^{-} \;=\; - \left(\chi -\frac{\alpha}{4}\right) M\,, 
\label{eqs:coef_int}
\eeqa

\noindent
where

\beq 
M_{\pm} \;=\; \int\dmath x_{\pm}\dmath y_{\pm}\,f^{\pm}(x_{\pm}, y_{\pm}, t)
\;=\; \mbox{constant} 
\label{masspm}
\eeq 

\noindent
are the (constant) masses in $\pm$ populations, and $M = M_{+} + M_{-}$ is the total mass in both discs. 

The coefficients $\left(A_{\pm}, B_{\pm}, C_{\pm}, D_{\pm}, E_{\pm}, F_{\pm}, G_{\pm}\right)$ are all, in general, functions of time, whereas $K$ is a constant proportional to the total mass in both discs. Since we have defined separate canonical variables for the $\pm$ populations, it is necessary to take care to write the CBEs of equation~(\ref{cbepm}) as 
\beq
\frac{\partial f^{+}}{\partial t} \;+\; \left[f^{+}, H^{+} \right]_{+} \;=\; 0\,; \qquad\quad
\frac{\partial f^{-}}{\partial t} \;+\; \left[f^{-}, H^{-} \right]_{-} \;=\; 0\,,
\label{cbepmsep}
\eeq

\noindent
where we have put $\pm$ subscripts on the Poisson Brackets to indicate that they are to be taken with respect to the appropriate set of canonical variables. Then the CBEs in equation~(\ref{cbepmsep}) together with the expressions for $H^{\pm}$  given in equations~(\ref{hampm}) and (\ref{eqs:coef_int}) completely define the self--consistent evolution of the counter--rotating discs, accurate to $4^{\rm th}$ order in the eccentricities. 

\section{Counter--rotating discs: time--dependent DFs}

In galactic dynamics, it is possible to construct many steady state solutions 
of the self--consistent CBE, whereas time--dependent behaviour is very difficult to understand even in the linearized limit. However, the present case of counter--rotating discs of small eccentricities around a central object turns out to be more tractable. The self--consistent dynamics described by equations~(\ref{hampm})---(\ref{cbepmsep}) has implicit in it a certain approximate integrable dynamics. This remarkable circumstance allows us to construct approximate time--dependent, self--consistent DFs, and describe the evolution of the counter--rotating instability largely analytically. 

\subsection{An approximate dynamical invariant}

In this subsection we are interested in the construction of an approximate invariant for the dynamics on a two dimensional phase space, generated by a
time--dependent Hamiltonian which is similar in form to those of 
equations~(\ref{hampm}). We drop the $\pm$ signifiers on all quantities in the interests of reducing clutter, but will restore them in the next subsection. Hence consider the Hamiltonian

\beqa
H(x, y, t) &=& \frac{1}{2}A(t)x^2 \;+\; B(t)xy \;+\; \frac{1}{2}C(t)y^2 
\;+\; D(t)x \;+\; E(t)y \nonumber\\[1em]
&& +\, F(t)x\left(x^2 + y^2\right) \;+\; G(t)y\left(x^2 + y^2\right) \;+\;
K\left(x^2 + y^2\right)^2\,.
\label{hamh}
\eeqa

\noindent
This is a time--dependent Hamiltonian acting on the two--dimensional phase space $(x, y)$. We now seek to eliminate the linear terms in $H$ by making a canonical transformation to new coordinate and momentum, $\left({\xi}_{1}, {\xi}_{2}\right)$, through a generating function 

\beq
S(x, {\xi}_{2}, t) \;=\; \left[x \,-\, X(t)\right]\left[{\xi}_{2} \,+\, Y(t)\right]\,,
\label{sdef}
\eeq

\noindent
where $X(t)$ and $Y(t)$ are some time--dependent functions, which are to be determined. Since $y = \left(\partial S/\partial x\right)$ and $\xi_1 = \left(\partial S/\partial \xi_2\right)$, the transformation
 
\beq
x \;=\; {\xi}_{1} + X(t)\,;\qquad\quad y \;=\; {\xi}_{2} + Y(t)\,,
\label{xytrans}
\eeq

\noindent
amounts to a time--dependent shift of the origin of phase space. The new Hamiltonian is given by

\beqa
H_{\rm t}\left({\xi}_{1}, {\xi}_{2}, t\right) &=& H\left(x\left({\xi}_{1}, {\xi}_{2}\right), y\left({\xi}_{1}, {\xi}_{2}\right), t\right) \;+\; \frac{\partial S}{\partial t}\nonumber\\[1em]
&=& H_{\rm lin} \;+\; H^{(0)}_{\rm t} \;+\; H^{(1)}_{\rm t}\,,
\label{newham}
\eeqa

\noindent
where $H_{\rm lin}$, $H^{(0)}_{\rm t}$ and $H^{(1)}_{\rm t}$ contain terms that are linear, quadratic, and cubic plus fourth order in $\left({\xi}_{1}, {\xi}_{2}\right)$, respectively. It is straightforward to work out that

\beqa
H_{\rm lin} &=& A X {\xi}_{1} \;+\; B Y {\xi}_{1} \;+\; B X {\xi}_{2} \;+\; C Y {\xi}_{2} \;+\; D {\xi}_{1} \;+\; E {\xi}_{2}\nonumber\\[1em]
&+& F {\xi}_{1} (X^2 + Y^2) \;+\; 2 F X ( X {\xi}_{1} \;+\;  Y {\xi}_{2}) 
\;+\; G {\xi}_{2} (X^2 + Y^2) \;+\; 2 G Y (X {\xi}_{1} + Y {\xi}_{2}) \nonumber \\[1em]
&+& 4 K [(Y^3+ Y X^2){\xi}_{2} + (X^3+X Y^2){\xi}_{1}] \;+\; \frac{\dmath Y}{\dmath t} {\xi}_{1} \;-\; \frac{\dmath X}{\dmath t}{\xi}_{2}\,.
\label{hlin} 
\eeqa

\noindent
We now require that $H_{\rm lin}$ vanishes. This happens when $X(t)$ and $Y(t)$ obey the following first order ordinary differential equations (ODEs):

\beqa
\frac{\dmath X}{\dmath t} &=& B X \;+\; C Y \;+\; E \;+\; 2 FX Y \;+\; G(X^2 + 3 Y^2) \;+\; 4 K (Y^3 + Y X^2) \nonumber \\[1em]
\frac{\dmath Y}{\dmath t} &=& -\,A X \;-\; B Y \;-\; D \;-\; F (3 X^2 + Y^2) \;-\; 2 G X Y \;-\; 4 K (X^3 + X Y^2)\,.
\label{eqs:xy}
\eeqa

Having eliminated $H_{\rm lin}$, the remaining terms in the new Hamiltonian are $H^{(0)}_{\rm t}$ and $H^{(1)}_{\rm t}$. We require the former:\footnote{The last bit of the new Hamiltonian, $H^{(1)}_{\rm t} =  F_{\rm t}\xi_1\left(\xi_1^2 + \xi_2^2\right) + G_{\rm t}\xi_2\left(\xi_1^2 + \xi_2^2\right) + K\left(\xi_1^2 + \xi_2^2\right)^2\,$, contains cubic and fourth order terms in $\left(\xi_1, \xi_2\right)\,$. The new coefficients, $F_{\rm t} = F + 4KX\,$ and $G_{\rm t} = G + 4KY\,$, are given in terms of the old coefficients and the centroid coordinates. Henceforth we will not consider the modification of the dynamics due to these higher order terms.}    

\beq
H_{\rm t}^{(0)}\left({\xi}_{1}, {\xi}_{2}, t\right) \;=\; \frac{1}{2} A_{\rm t}(t) {\xi}_{1}^{2} \;+\; B_{\rm t}(t) {\xi}_{1} {\xi}_{2} \;+\; \frac{1}{2} C_{\rm t}(t) {\xi}_{2}^{2}\,,
\label{ht0def} 
\eeq

\noindent
where the new coefficients,

\beqa
\frac{A_{\rm t}}{2} &=& \frac{A}{2} \;+\; 3 F X \;+\; G Y \;+\; 2 K(Y^2 + 3 X^2) \nonumber \\[1em]
B_{\rm t} &=& B \;+\; 2 F Y \;+\; 2 G X \;+\; 8 K X Y \nonumber\\[1em]
\frac{C_{\rm t}}{2} &=& \frac{C}{2} \;+\; F X \;+\;  3 G Y \;+\; 2 K(X^2 + 3 Y^2)\,,
\label{newcoeff}
\eeqa

\noindent
are given in terms of the old coefficients and the centroid coordinates.

The homogeneous, linear and time--dependent dynamics generated by
$H_{\rm t}^{0}$ preserves areas in $\left({\xi}_{1}, {\xi}_{2}\right)$
space. Moreover, initial conditions given on any ellipse centered at
$\xi_1=0$ and $\xi_2=0$ will, at a later time, lie on some other
centered ellipse of the same area.\footnote{The shape --- i.e. axis
ratio --- and orientation are determined by the matrix ODE
Eq.~\ref{eqs:ellipse}.} Therefore, there must be a quadratic quantity
that is preserved by the dynamics. Let us write this invariant as:
 
\beq
J \;=\; \frac{1}{2}\bbxi^{\rm T}\,\bfQ(t)\,\bbxi \;=\; \frac{1}{2} Q_{i\,j}(t) {\xi}_{i} {\xi}_{j}\,,
\label{invj}
\eeq 

\noindent
where $\bfQ(t)$ is a time--dependent, positive definite, $2 \times 2$ matrix. Because phase areas are conserved, ${\det} (\bfQ)$ is constant. The linear dynamics generated by $H_{\rm t}^{(0)}$ can be written in matrix form as:
\beq
\frac{\dmath\bbxi}{\dmath t} \;=\; \bfT(t)\bbxi\,;\qquad\quad
\bfT(t) \;=\; \left(\bea{lr} B_{\rm t} & C_{\rm t} \\ 
-A_{\rm t} & - B_{\rm t}\eea\right)\,,
\label{eomhto}
\eeq
\noindent
where we have introduced $\bfT(t)$, which is a time--dependent, traceless,  
$2 \times 2$ matrix. If $J$, given in equation~(\ref{invj}), is an invariant of the dynamics, we must have
\beq
\frac{\dmath J}{\dmath t} \;\equiv\; \frac{1}{2}\bbxi^{\rm T} \left[\bfT^{\rm T} \bfQ \,+\,  \bfQ  \bfT \,+\, \frac{d\bfQ}{dt} \right]\bbxi \;=\; 0\,. 
\label{jdotzero}
\eeq 
\noindent
Therefore $\bfQ(t)$ must obey the matrix ODE:
\beq  
\frac{\dmath\bfQ}{\dmath t} \;=\; -\,\bfT^{\rm T} \bfQ \;-\; \bfQ\,\bfT\,.
\label{eqn:Q}
\eeq
\noindent
It can be verified that the equation above preserves ${\det}(\bfQ)$.

\subsection{Distribution functions}
\label{sec:df}
We are now ready to deal with the self--consistent dynamics of counter--rotating discs, described by equations~(\ref{hampm})---(\ref{cbepmsep}).
We now restore the $\pm$ signs that were dropped in the previous subsection.
The first step is to pass from the Hamiltonians $H^{\pm}$ of equation~(\ref{hampm}) to new $\pm$ Hamiltonians $H_{\rm t}^{(0)\pm}$ which are of the form given by equation~(\ref{ht0def}). We do not need to write down these new Hamiltonians; it suffices to note that they possess quadratic, time--dependent invariants of the form given in equations~(\ref{invj}): 

\beqa
J_{\pm}(x_{\pm}, y_{\pm}, t) &\;=\;& \frac{1}{2} Q^{\pm}_{11}(t)\,\left[x_{\pm} - X_{\pm}(t)\right]^{2} \,\;+\;\,
Q^{\pm}_{12}(t)\,\left[x_{\pm} - X_{\pm}(t)\right]\left[y_{\pm} - Y_{\pm}(t)\right] \,\nonumber\\[1em]
&&\qquad\qquad +\;\;\frac{1}{2} Q^{\pm}_{22}(t)\left[y_{\pm} - Y_{\pm}(t)\right]^{2}\,.
\label{jpmdef}
\eeqa

\noindent

\begin{figure}
  \centering
  \begin{minipage}{1.0\linewidth}
    \centering
    \includegraphics[width=3.0in]{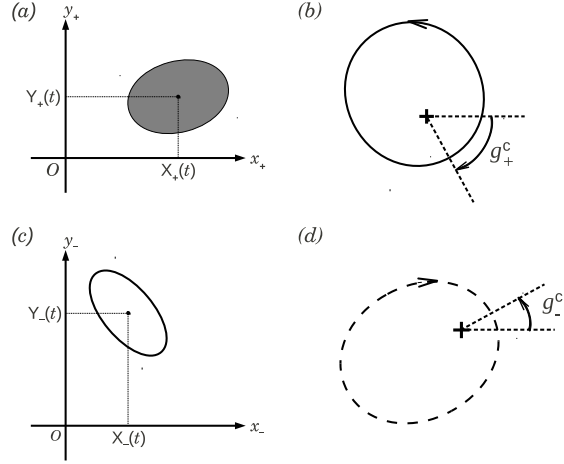}  
  \end{minipage}

\caption[Distributions]{Distribution functions and centroid orbits:
Panels "a" and "c" are schematic illustrations of the $\pm$ stellar
distributions in the $\left(x_{\pm}, y_{\pm}\right)$ phase spaces,
respectively, at some given time. The elliptical patches indicate the
regions populated by stars; each point in this phase space corresponds
to a Keplerian orbit in physical space.  The centroids of the
ellipses, $\left(X_{\pm}(t),Y_{\pm}(t)\right)$, are marked by black
dots, and the principal axes of the ellipses are given by the
eigenvectors of the symmetric matrices, ${\bf Q}^\pm(t)\,$. Panels "b"
and "d" are representations, in physical space, of the Keplerian
orbits corresponding to the $\pm$ centroids. Note that the angles to
the pericentre, $g_{\pm}^c = \mp\,\rm arctan{(X_{\pm}/Y_{\pm})}\,$;
hence, for the distributions shown in panels "a" and "c", we have
$g_+^c < 0\,$ and $g_-^c > 0\,$.}  \label{fig:distributions}
\end{figure} The DFs $f^{\pm}(x_{\pm}, y_{\pm}, t)$ obey the CBEs of
equations~(\ref{cbepmsep}). By Jeans' theorem \citep{bt08}, any
function of the dynamical invariants is a solution of the
CBE. Therefore we choose the $\pm$ DFs to be functions of the
approximate invariants $J_{\pm}\,$:

\beq
f^{\pm}(x_{\pm}, y_{\pm}, t) \;=\; F^{\pm}\left(J_{\pm}\right)\,.
\label{dfchoice}
\eeq

\noindent
Schematic representations of DFs in the $\left(x_{\pm}, y_{\pm}\right)$ phase-spaces, and of centroid orbits in physical space, are shown in Fig.(\ref{fig:distributions}). 
Here, we note the following important properties:

\begin{enumerate}
\item The DFs $F^{\pm}(J_{\pm})$ are assumed to have compact support over
the interval $0 \leq J_\pm \leq J_{\pm \rm max} \ll 1\,$.

\item $X_\pm(t)$ and $Y_\pm(t)$ are the coordinates of the {\em centroids} of the DFs, $\,F^{\pm}\left(J_{\pm}\right)\,,$ in the $\left(x_\pm, y_\pm\right)$ phase spaces, respectively.

\item At any instant of time, the isocontours of the DFs are ellipses in the 
$\left(x_\pm, y_\pm\right)$ phase spaces that are centered on $\left(X_\pm(t), Y_\pm(t)\right)$. 

\item The shapes and orientations of the ellipses are described by the time--dependent, symmetric, positive definite matrices $\bfQ^{\pm}(t)$, which we will refer to as the {\em shape matrices}. Time evolution preserves ${\det}\left(\bfQ^{\pm}\right)$; we can choose ${\det}\left(\bfQ^{\pm}\right) = 1\,$.

\item The zeroth and first moments of the DFs,
\beq
M_{\pm} \;=\; {\int}_{0}^{J_{\pm \rm max}} 2 \pi dJ_{\pm} \,F^{\pm}(J_{\pm})\,; \qquad\quad
{\sigma}^{2}_{\pm} \;=\; \frac{1}{M_{\pm}}{\int}_{0}^{J_{\pm \rm max}} 2 \pi dJ_{\pm}\,J_{\pm}\, F^{\pm}(J_{\pm})\,,
\label{msigma}
\eeq

\noindent
are the disc masses and the squared dispersions (of eccentricities), respectively. Once the DFs have been specified, $M_\pm$ and $\sigma^2_\pm$ can be treated as constants.

\item
We can now state precisely the conditions under which $F^{\pm}(J_{\pm})$
are approximate solutions. These DFs are good solutions when the dispersions
in the eccentricities are much smaller than their centroid values; i.e. when

\beq
0 \;\leq\; {\sigma}_{\pm} \;\ll\; e_\pm \;\ll\; 1\,,
\label{cond}
\eeq

\noindent
where $e_\pm$ is a typical value of $\sqrt{X^2_\pm + Y^2_\pm}\,$.

\item The total angular momentum in the two discs is
\beq
\mathcal{L}_{\rm tot} \;=\; \sqrt{G\mbul a_0}\int\dmath \ell\dmath g\; \ell \,f^+ \;\;+\;\;\sqrt{G\mbul a_0}\int\dmath \ell\dmath g\; \ell\, f^- \,,
\label{angmomdef}
\eeq
\noindent where $\ell$ is positive for $f^+$ with $\dmath \ell \dmath g = \dmath x_{+} \dmath y_{+}$, and $\ell$ is negative for $f^-\,$ with $\dmath \ell \dmath g = \dmath x_{-} \dmath y_{-}$. Using equations~(\ref{ithetadef})---(\ref{xyminus}) and (\ref{xytrans}), we write 
\beq
\ell \;=\; \pm\left[1 \;-\; \frac{1}{2}\left(x_\pm^2 \,+\, y_\pm^2\right)\right]
\;=\; \pm\left[1 \;-\; \frac{1}{2}\left\{\left(\xi_{1\pm} \,+\, X_\pm\right)^2 \;+\; \left(\xi_{2\pm} \,+\, Y_\pm\right)^2\right\}\right]  
\label{angmomexp}
\eeq

\noindent
For the DFs given in equation~(\ref{dfchoice}),  $f^{\pm}(x_{\pm}, y_{\pm}, t) = F^{\pm}\left(J_{\pm}\right)$. When equation~(\ref{angmomexp}) is used, 
$\xi_{1\pm}$ and $\xi_{2\pm}$ contribute to the integral of equation~(\ref{angmomdef}) only at second order. These contributions are small, since 
the dispersion in eccentricities is much smaller than centroid values for these DFs. So we drop the dependences on $\xi_{1\pm}$ and $\xi_{2\pm}$ on the right side of equation~(\ref{angmomexp}), and set $x_\pm = X_\pm$ and $y_\pm = Y_\pm\,$. Using the definitions of $M_\pm$ given in the first of equations~(\ref{msigma}) above, we obtain

\beq
\frac{\mathcal{L}_{\rm tot}}{\sqrt{G\mbul a_0}} \;=\; \left[M_+ \,-\, M_-\right] \;-\; \frac{M_+}{2}\left[X_+^2 \,+\, Y_+^2\right] 
\;+\; \frac{M_-}{2}\left[X_-^2 \,+\, Y_-^2\right] \;+\; O\left({\sigma}_{\pm}^2/e_\pm^2\right)\,.
\label{angmomfin} 
\eeq
\noindent The first term on the right side is the contribution from the $\pm$ discs if they were circular; the second term is the decrement due to the centroid eccentricity of the $+$ disc; the third term is a similar and oppositely signed contribution from the $-$ disc.
\end{enumerate}

We need to compute the coefficients $\left(A_{\pm}, B_{\pm}, C_{\pm}, D_{\pm}, E_{\pm}, F_{\pm}, G_{\pm}, K_{\pm}\,\right)$, by substituting equation~(\ref{dfchoice}) for the $\pm$ DFs in equations~(\ref{eqs:coef_int}). Similar to the treatment of the angular momentum of the discs given above, we set $x_\pm = X_\pm$ and $y_\pm = Y_\pm\,$. Then, it is straightforward to express the coefficients as functions of the centroid coordinates, $\left[X_{\pm}(t)\,,\, Y_{\pm}(t)\right]$:

\beqa
 A_{\pm} &=& -\,2 M\alpha \;-\; 2M_{\pm}\left[(\gamma + \lambda){X}_{\pm}^2 + \gamma {Y}_{\pm}^2 \right] \;-\; 2 M_{\mp}\left[(\gamma + \lambda){X}_{\mp}^2 + \gamma {Y}_{\mp}^2 \right]\nonumber\\[2em]
 B_{\pm} &=&  -\,2 \lambda M_{\pm}{X}_{\pm}{Y}_{\pm}  \;+\; 2 \lambda M_{\mp}{X}_{\mp}{Y}_{\mp}\nonumber \\[2em]
C_{\pm} &=& -\,2 M\alpha \;-\; 2M_{\pm}\left[\gamma {X}_{\pm}^2 + (\gamma + \lambda){Y}_{\pm}^2  \right] \;-\; 2 M_{\mp}\left[\gamma {X}_{\mp}^2 + (\gamma + \lambda){Y}_{\mp}^2\right]\nonumber\\[2em]
D_{\pm} &=& -\beta M_{\pm} X_{\pm} \,-\, \left(\kappa -\frac{\beta}{8}\right)M_{\pm} \left[{X}_{\pm}^3  + X_{\pm}Y_{\pm}^{2}\right] \,+\,
\beta M_{\mp} X_{\mp} \,+\, \left(\kappa -\frac{\beta}{8}\right) M_{\mp}\left[{X}_{\mp}^3  + X_{\mp}Y_{\mp}^{2} \right]\nonumber\\[2em]
E_{\pm} &=& -\,\beta M_{\pm} Y_{\pm} \;-\; \left(\kappa -\frac{\beta}{8}\right)M_{\pm} \left[ {Y}_{\pm}^3 + X_{\pm}^{2}Y_{\pm}\right] \;-\;
\beta M_{\mp} Y_{\mp} \;-\; \left(\kappa -\frac{\beta}{8}\right)M_{\mp} \left[ {Y}_{\mp}^3 + X_{\mp}^{2}Y_{\mp}\right]\nonumber\\[2em]
F_{\pm} &=& -\left(\kappa - \frac{\beta}{8}\right)\left[M_{\pm}X_{\pm} - M_{\mp}X_{\mp}\right]\nonumber\\[1em]
G_{\pm} &=& -\left(\kappa - \frac{\beta}{8}\right)\left[M_{\pm}Y_{\pm} + M_{\mp}Y_{\mp}\right]\nonumber\\[1em]
K_{\pm} &=& -\left(\chi - \frac{\alpha}{4}\right)M\,.
\label{eqs:coef_ellipse}
\eeqa

\noindent
The centroid coordinates obey:

\beqa
\frac{\dmath X_\pm}{\dmath t} &=& B_\pm X_\pm \;+\; C_\pm Y_\pm \;+\; E_\pm \;+\; 2 F_\pm X_\pm Y_\pm \;+\; G_\pm\left(X_\pm^2 + 3 Y_\pm^2\right) \;+\; 4 K_\pm\left(Y_\pm^3 + Y_\pm X_\pm^2\right)\nonumber \\[1em]
\frac{\dmath Y_\pm}{\dmath t} &=& -A_\pm X_\pm - B_\pm Y_\pm - D_\pm - F_\pm\left(3 X_\pm^2 + Y_\pm^2\right) - 2 G_\pm X_\pm Y_\pm - 4 K_\pm \left(X_\pm^3 + X_\pm Y_\pm^2\right)\,.\nonumber\\[1ex]
&&\label{eqs:xypm}
\eeqa

\noindent
These are a set of 4 autonomous first order ODEs with cubic nonlinearity.
The shape matrices obey the following first order matrix ODEs:

\beq  
\frac{d\bfQ^\pm}{dt} \;=\; -\,\left(\bfT^\pm\right)^{\rm T} \bfQ^\pm \;-\; \bfQ^\pm\,\bfT^\pm\,;\qquad\quad
\bfT^\pm(t) \;=\; \left(\bea{lr} \;B_{\rm t\pm} & C_{\rm t\pm} \\[2em] 
-\,A_{\rm t\pm} & \;-\,B_{\rm t\pm}\eea\right)\,,
\label{eqs:ellipse}
\eeq

\noindent
where 

\beqa
A_{\rm t\pm} &=& A_\pm \;+\; 6 F_\pm X_\pm \;+\; 2G_\pm Y_\pm \;+\; 4 K_\pm\left(Y_\pm^2 + 3 X_\pm^2\right) \nonumber \\[1em]
B_{\rm t\pm} &=& B_\pm \;+\; 2 F_\pm Y_\pm \;+\; 2 G_\pm X_\pm \;+\; 8 K_\pm X_\pm Y_\pm \nonumber\\[1em]
C_{\rm t\pm} &=& C_\pm \;+\; 2F_\pm X_\pm \;+\;  6 G_\pm Y_\pm \;+\; 4 K_\pm\left(X_\pm^2 + 3 Y_\pm^2\right)\,.
\label{newcoeffpm}
\eeqa

\noindent
Since $\bfT^\pm(t)$ are traceless matrices, $\det\left[\bfQ^\pm(t)\right]$ is a conserved quantity; without loss of generality, we can choose $\det\left[\bfQ^\pm(0)\right] = 1\,$. The matrix $\bfT$ depends on the centroid coordinates, so the matrix equations for $\bfQ$, while linear, are driven by centroid evolution. Equations~(\ref{eqs:coef_ellipse})---(\ref{newcoeffpm}) determine the self--consistent centroid and shape dynamics of DFs describing the counter--rotating discs.

\section{Counter--rotating discs: centroid dynamics} 

As shown above, in the limit $0\,\leq\,{\sigma}_{\pm}\,\ll\, e_\pm
\,\ll\, 1\,,$ shape dynamics is driven by centroid dynamics, and
consists of area preserving evolution of the shape and orientation of
the elliptical isocontours of the DFs, with no feedback on
centroids.\footnote{Such a feedback requires a higher order theory,
and may very well account for instability saturation in a planar
analog of the three dimensional saturation described in
\citet{ttk09}.} In what follows, and with the understanding that shape
can be recovered easily from centroids as and when required by a given
application, we drop any further reference to shape dynamics, and
focus our attention on the nonlinear evolution of the
centroids.

\subsection{Integrability}
 
The centroid equations~(\ref{eqs:xypm}) are a set of 4
autonomous first order ODEs with cubic nonlinearity. Quite remarkably,
it turns out that they describe a non linear, yet integrable,
system. This happens because of the underlying Hamiltonian structure
and the presence of a second conserved quantity. Let us define new
rescaled variables $\left(u_\pm, v_\pm\right)$:

\beqa
u_{\pm} &=& \sqrt{{\mu}_{\pm}}\;X_{\pm}\,,\qquad v_{\pm} \;=\; \sqrt{{\mu}_{\pm}}\;Y_{\pm}\nonumber\\[1em]
{\mu}_{\pm} &=& \frac{M_{\pm}}{M}\,,\qquad\mbox{so}\quad \left({\mu}_{+} \,+\, {\mu}_{-}\right) \;=\; 1\,.
\eeqa

\noindent
A lengthy yet straightforward calculation shows that the centroid equations~(\ref{eqs:xypm}) are equivalent to the following two degree--of--freedom system Hamiltonian system, with coordinates $u_\pm$ and momenta $v_\pm\,$: 

\beqa
\frac{\dmath u_+}{\dmath t} &=& \frac{\partial {\mathcal H}}{\partial v_+}\,,\qquad\frac{\dmath v_+}{\dmath t} \;=\; -\frac{\partial {\mathcal H}}{\partial u_+}\,;\qquad\quad
\frac{\dmath u_-}{\dmath t} \;=\; \frac{\partial {\mathcal H}}{\partial v_-}\,,\qquad\frac{\dmath v_-}{\dmath t} \;=\; -\frac{\partial {\mathcal H}}{\partial u_-}\,,\nonumber\\[3em]
{\mathcal H} &=& -\left(\frac{\beta M {\mu}_{+}}{2} \,+\, M\alpha\right) \left[u_{+}^{2} \,+\, v_{+}^{2}\right] \;-\; \left(\frac{\beta M {\mu}_{-}}{2} \,+\, M\alpha\right) \left[u_{-}^{2} \,+\, v_{-}^{2}\right] \nonumber\\[2em]
&& +\;\beta M \sqrt{{\mu}_{+}{\mu}_{-}}\,\left[u_{+}u_{-} \,-\, v_{+}v_{-}\right] \nonumber\\[2em]
&& - \left(\frac{\gamma + \lambda}{2} \,+\, \frac{4\chi -\alpha}{4\mu_+} \,+\, \kappa \,-\, \frac{\beta}{8}\right)M \left[u_{+}^2 \,+\, v_{+}^2\right]^2 \nonumber\\[2em]
&& - \left(\frac{\gamma + \lambda}{2} \,+\, \frac{4\chi -\alpha}{4\mu_-} \,+\, \kappa \,-\, \frac{\beta}{8}\right)M \left[u_{-}^2 \,+\, v_{-}^2\right]^2 \nonumber\\[2em]
&& -\; \lambda M \left[u_{+} u_{-} \,-\, v_{+} v_{-}\right]^2 \;-\; \gamma M \left[\left(u_{+}^2+v_{+}^2\right) \left(u_{-}^2+v_{-}^2\right)\right] \nonumber\\[2em]
&& +\; \left(\kappa - \frac{\beta}{8}\right)M \left[u_{+} u_{-} \,-\, v_{+} v_{-}\right]\left[\sqrt{\frac{{\mu}_{+}}{{\mu}_{-}}} \left(u_{-}^2 \,+\, v_{-}^2\right) \,+\, \sqrt{\frac{{\mu}_{-}}{{\mu}_{+}}} \left(u_{+}^2 \,+\, v_{+}^2\right)\right]\,,
\label{eqn:uv}
\eeqa

\noindent
where ${\mathcal H}\left(u_+,u_-,v_+,v_+\right)$ is the Hamiltonian for the two degree--of--freedom system with coordinates $u_\pm$ and momenta $v_\pm\,$. Since ${\mathcal H}$ is independent of time, it is conserved. It is straightforward to verify that equations~(\ref{eqn:uv}) also conserve the total angular momentum defined in equation~(\ref{angmomfin}). Dropping the first term, we write the second conserved quantity as

\beq
{\mathcal L} \;=\; \frac{u_{+}^2 \,+\, v_{+}^2}{4} \;-\; \frac{u_{-}^2 \,+\, v_{-}^2}{4}\,.
\label{angmomdec}
\eeq

\noindent
This quantity is a measure of the amount by which the angular momentum is lower
than the maximum value it can attain when the centroid eccentricities are zero; for brevity we shall henceforth refer to ${\mathcal L}$ as the angular momentum.
 
Since this two degree--of--freedom system has two independent conserved quantities, ${\mathcal H}$ and ${\mathcal L}$, it is integrable. We will explore the non linear dynamics of this system, after examining the linear instability of zero eccentricity discs.

\subsection{Linear instability of zero eccentricity discs}
\label{sec:linear}
The zero eccentricity state, $u_{\pm} = v_{\pm} = 0$ which has ${\mathcal L} =0\,$, is an equilibrium state of the dynamics governed by ${\mathcal H}$.\footnote{Implicit to this zero eccentricity equilibrium are DFs of the form $F^{\pm}(J_{\pm})= \frac{M_{\pm}}{2\pi} \delta(J_{\pm})$, i.e, DFs with ${\sigma}^{2}_{\pm} = 0$ and no shape (${\bf Q}^{\pm}$) dynamics to speak of.} Here we determine the conditions under which this equilibrium is unstable to small perturbations. The linearized equations obeyed by infinitesimal perturbations are:

\beqa
\frac{\dmath u_+}{\dmath t} &=& - w_{+} v_{+} \;-\; w_c v_{-}\,,\qquad\quad
\frac{\dmath v_+}{\dmath t} \;=\; w_{+} u_{+} \;-\; w_c u_{-}\,;\nonumber\\[2em]
\frac{\dmath u_-}{\dmath t} &=& - w_{-} v_{-} \;-\; w_c v_{+}\,,\qquad\quad
\frac{\dmath v_-}{\dmath t} \;=\; w_{-} u_{-} \;-\; w_c u_{+}\,,
\label{uveqns}
\eeqa

\noindent
where $w_{\pm} = \left(2 M \alpha \,+\, \beta M_{\pm}\right)\,$, and $w_c = \beta \sqrt{M_{+}M_{-}}\;$ are constants. It is readily verified that this linearized system conserves ${\mathcal L}$. To solve these equations let us define the complex variables:

\beq
z_{+} \;=\; u_{+} \;+\; \im\, v_{+}\,,\qquad\quad z_{-} \;=\; v_{-} \;+\; \im\, u_{-}\,,
\label{zpmdef} 
\eeq
\noindent 
in terms of which equations~(\ref{uveqns}) reduce to:

\beq
\frac{\dmath z_+}{\dmath t} \;=\; \im\, w_{+} z_{+} \;-\; w_c z_{-}\,,\qquad\quad
\frac{\dmath z_-}{\dmath t} \;=\; -\im\, w_{-} z_{-} \;-\; w_c z_{+}\,.
\label{zeqns}
\eeq
\noindent
Note that the asymmetry in Eqs.~(\ref{zeqns}) is inherited from the asymmetry in the definition of $z_{\pm}$ in Eqs.~(\ref{zpmdef}). Looking for normal modes, $z_{\pm} = Z_{\pm} \e^{s t}\,$, we obtain
and solve a quadratic characteristic equation for the two eigenvalues:
\beq
s \;=\; \frac{\im}{2}\left({w_{+} - w_{-}}\right) \;\pm\; \frac{1}{2}\sqrt{4\,w_c^{2} \,-\, \left(w_{+} + w_{-}\right)^2}\;\;.
\label{eigen}
\eeq
\noindent
There is a growing solution when  $4w_c^2 > \left(w_{+} + w_{-}\right)^2$. Thus the zero eccentricity equilibrium, $u_{\pm} = v_{\pm} = 0\,$, is unstable (overstable) when

\beq
{\mu_{+}}{\mu_{-}} \;>\; \frac{1}{4} {\left(1+ \frac{4 \alpha}{\beta}\right)}^{2}\,.
\label{unstable}
\eeq

\begin{figure}
  \centering
  \begin{minipage}{1.0\linewidth}
    \centering
    \includegraphics[width=3.0in]{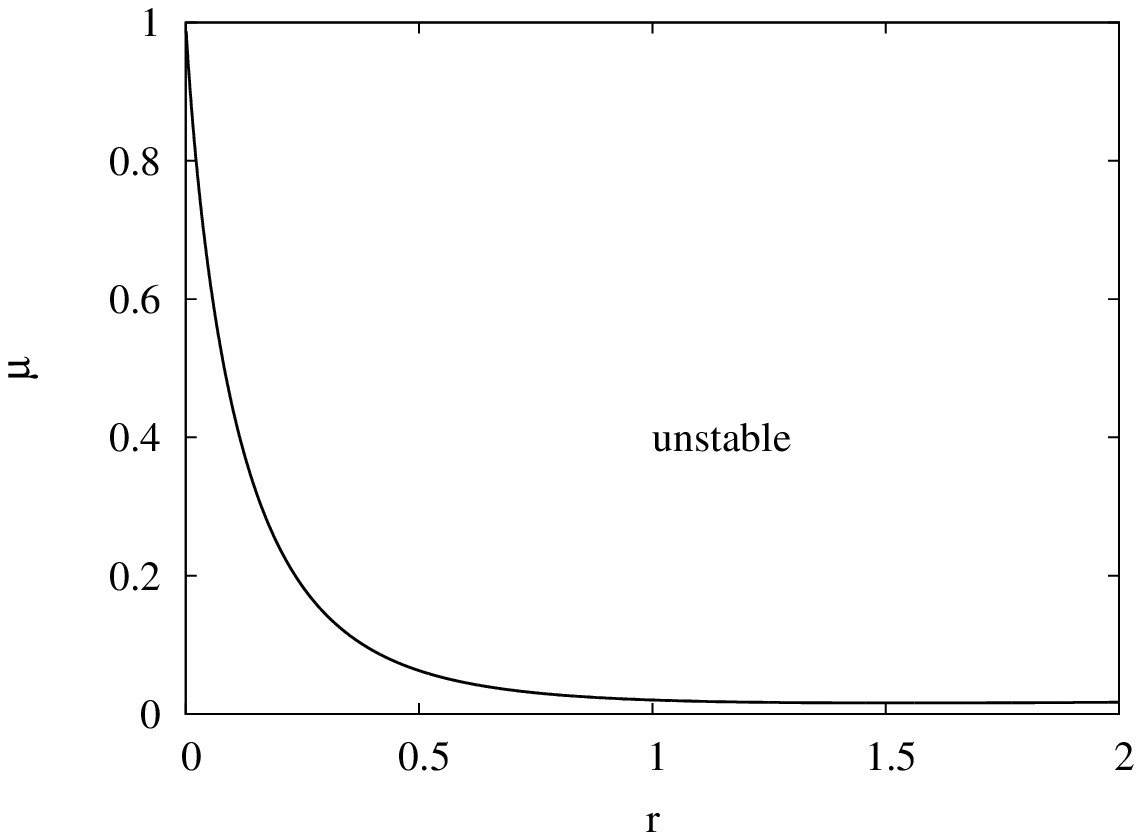}  
    \includegraphics[width=3.0in]{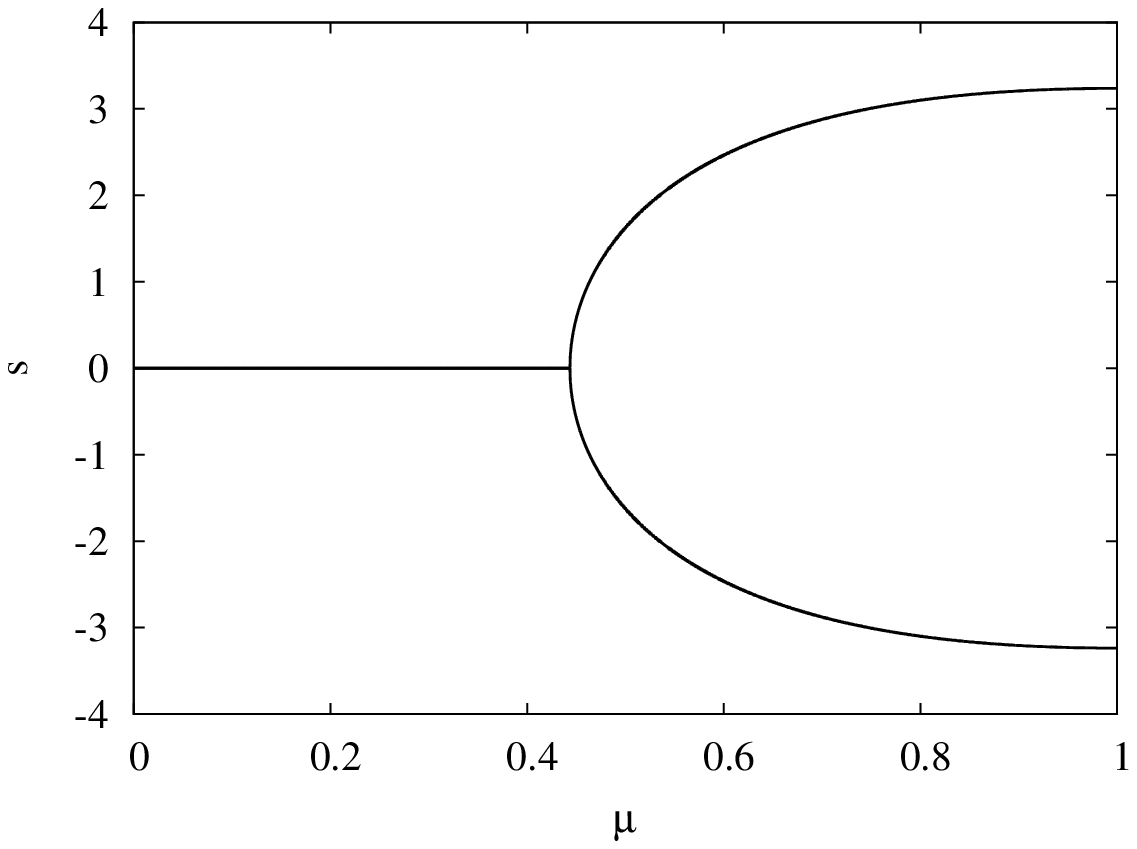}  
  \end{minipage}

\caption[Stability]{Linear Stability of the zero angular momentum equilibrium: {\em Left}: Curve of critical values in the plane of mass ratio $\mu$ and dimensionless softening $r=b/a_0$ above which the centroids of the prograde and retrograde population are unstable to eccentricity growth. {\em Right}: The real parts of the eigenvalues given in 
equations~(\ref{eigen}), as a function of $\mu$ for $r=0.1\,$. Instability sets in for $\mu \geq \mu_{\rm crit} \simeq 0.4436\,$.}
\label{fig:stability}
\end{figure}

\noindent Using the definitions of $\alpha$ and $\beta$ given in the
Appendix~A, we display the stability criterion of
equation~(\ref{unstable}) in Fig.(\ref{fig:stability}) by plotting the
mass ratio, $\mu = M_{-}/M_{+}\,$, versus the dimensionless softening,
$r = b/a_0\,$.\footnote{Without loss of generality, we assume that
$0\leq\mu\leq 1\,$.} For a given softening, there is a critical value
of the mass ratio (which decreases with increasing softening) above
which the zero--eccentricity equilibrium is unstable. Conversely, for
a given mass ratio, there is a critical value of the softening (which
increases with decreasing mass ratio) above which the
zero--eccentricity equilibrium is unstable. In the right panel of
Fig.(\ref{fig:stability}) we plot the real parts of the eigenvalues
given in equations~(\ref{eigen}), as a function of $\mu$ for
$r=0.1\,$. When $\mu \leq \mu_{\rm crit} \simeq 0.4436\,$, the
eigenvalues are both imaginary, corresponding to normal modes
describing steadily precessing discs of fixed centroid
eccentricities. Both growing and damped solutions are allowed when
$\mu > \mu_{\rm crit}\,$. The growing (damped) solution describes
discs whose eccentricities grow (damp) as they precess steadily.

Since ${\mathcal L}$ is conserved, the instability operates through
exchange of angular momentum between the prograde and retrograde
discs. When the prograde disc gives some angular momentum to the
retrograde disc, both discs increase their eccentricities.  As is
well--known \citep{mil71, bt08}, the softening length mimics the
epicyclic radius of stars on nearly circular orbits.\footnote{Our
  general formalism does not of course need softening. But, the DFs
  considered in Sec.\ref{sec:df} and after are cold in the sense that
  the dispersions in the eccentricities are much smaller than the
  centroid eccentricities. So, softening serves to mimic eccentricity
  dispersion in our model (just like it mimics velocity dispersion in
  disc dynamics).} For this exchange to be self-reinforcing, for the
  instability to kick in, the mass ratio has to be large enough for
  $\pm$ disk self-gravity to overcome the effective ``heat'' due to
  softening (a process which is similar to the one driving the
  radial orbit instability \citep{lb1979}).

\subsection{Nonlinear dynamics}

As discussed earlier centroid dynamics is integrable, because it
is a two degree--of--freedom system (4--dimensional phase space) with
two independent conserved quantities. We now use the conservation of
the angular momentum of equation~(\ref{angmomdec}) to convert the
problem into a Hamiltonian system of one degree--of--freedom
system. We achieve this through two canonical transformations to new
variables. First, we pass from $\left(u_\pm, v_\pm\right)$ to new
action--angle variables, $\left(L_\pm, \psi_\pm\right)$:

\beqa
u_{+} &=& \sqrt{2 {L}_{+}}\,\sin{\psi}_{+}\,,\qquad\quad v_{+} \;=\; \sqrt{2 {L}_{+}}\,\cos{\psi}_{+}\;;\nonumber\\[1em]
u_{-} &=& \sqrt{2 {L}_{-}}\,\sin{\psi}_{-}\,,\qquad\quad v_{-} \;=\; \sqrt{2 {L}_{-}}\,\cos{\psi}_{-}\;.
\label{lpsidef}
\eeqa

\noindent
Written in terms of the new variables, the Hamiltonian of equations~(\ref{eqn:uv}) becomes:
\beqa
{\mathcal H} &=& -\,w_{+} {L}_{+} \;-\; w_{-} {L}_{-} \;-\; 2 w_c \sqrt{{L}_{+}{L}_{-}}\,\cos\left({\psi}_{+} \,+\, {\psi}_{-}\right) \nonumber\\[2em]
&& -\, 4{\eta}_{+} {L}_{+}^{2} \;-\; 4 {\eta}_{-} {L}_{-}^{2} \;-\; 4 \left(\kappa - \frac{\beta}{8}\right)M\sqrt{{L}_{+}{L}_{-}}\,\cos\left({\psi}_{+}+ {\psi}_{-}\right) \left[\sqrt{\frac{{\mu}_{+}}{{\mu}_{-}}}\,{L}_{-} \,+\, \sqrt{\frac{{\mu}_{-}}{{\mu}_{+}}}\,{L}_{+} \right] \nonumber \\[2em]
&& -\, 4\lambda M {L}_{+} {L}_{-}\,{\cos}^{2}\left({\psi}_{+}+ {\psi}_{-}\right) \;-\; 4 \gamma M {L}_{+} {L}_{-}\,,
\label{hamlpsi}
\eeqa

\noindent
where the new constants, $\eta_\pm$, are defined by:
 
\beq 
{\eta}_{\pm} \;=\; \left(\frac{\gamma + \lambda}{2} \,+\, \frac{4\chi -\alpha}{4\mu_\pm} \,+\, \kappa \,-\, \frac{\beta}{8}\right)M\,.
\label{etapm}
\eeq

\noindent

The angles ${\psi}_{+}$ and ${\psi}_{-}$ appear only in the combination $\left({\psi}_{+} + {\psi}_{-}\right)$. So we transform to new action--angle variables, $\left(\Sigma, \Theta\right)$ and $\left({\mathcal L}, \vartheta\right)$, defined through the generating function, ${\mathcal S}\left({\mathcal L}, \Sigma , {\psi}_{+} , {\psi}_{-}\right) = \left({\psi}_{+} + {\psi}_{-}\right)\Sigma \,+\, \left({\psi}_{+} - {\psi}_{-}\right){\mathcal L}\,$. Then,

\beqa
{L}_{+} &=& \Sigma \;+\; {\mathcal L}\,,\qquad\quad {L}_{-} \;=\; \Sigma \;-\; {\mathcal L}\,;\nonumber\\[1em]
\vartheta &=& {\psi}_{+} \;-\; {\psi}_{-}\,,\qquad\quad
\Theta \;=\; {\psi}_{+} \;+\; {\psi}_{-}\,.
\label{trsigthe}
\eeqa

\noindent
When expressed in terms of the new variables, the Hamiltonian becomes:

\beqa
{\mathcal H} &=& -\,\left(w_{+} + w_{-}\right)\Sigma \;-\; \left(w_{+} - w_{-}\right){\mathcal L} \;-\; 2 w_c \sqrt{{\Sigma}^2 \,-\, {\mathcal L}^2}\,\cos\Theta \nonumber\\[2em] 
&& -\, 4\left({\eta}_{+} + {\eta}_{-}\right){\Sigma}^2 \;-\; 8({\eta}_{+} - {\eta}_{-})\;{\Sigma} {\mathcal L} \;-\; 4\left({\eta}_{+} + {\eta}_{-}\right){\mathcal L}^2 \nonumber\\[2em]
&& -\, 4\left(\kappa - \frac{\beta}{8}\right) M \sqrt{{\Sigma}^2 - {\mathcal L}^2}\,\left[ \left(\sqrt{\frac{\mu_-}{\mu_+}} + 
\sqrt{\frac{\mu_+}{\mu_-}}\right)\Sigma \,+\, \left(\sqrt{\frac{{\mu}_{-}}{{\mu}_{+}}} - \sqrt{\frac{{\mu}_{+}}{{\mu}_{-}}}\right){\mathcal L} \right] \,\cos\Theta \nonumber\\[2em]
&& -\, 4\gamma M \left({\Sigma}^2 \,-\, {\mathcal L}^2\right) \;-\; 4\lambda M \left({\Sigma}^2 - {\mathcal L}^2\right)\,{\cos}^{2}\Theta\,.
\label{hamsigthe}
\eeqa

\noindent
Since ${\mathcal H}$ is independent of $\vartheta$, its conjugate momentum ${\mathcal L}$ is conserved.\footnote{From equations~(\ref{trsigthe}) and (\ref{lpsidef}), we have ${\mathcal L} = \left({L}_{+} - {L}_{-}\right)/2 = \left(u_{+}^2 + v_{+}^2 - u_{-}^2 - v_{-}^2\right)/4$ equal to the conserved quantity defined earlier in equation~(\ref{angmomdec}).} Hence ${\mathcal L}$ may be treated as a constant parameter occurring in ${\mathcal H}$, which can be thought of as a Hamiltonian describing the one degree--of--freedom Hamiltonian with coordinate $\Theta$ and momentum $\Sigma$. 

The global structure of dynamics in the $\left(\Theta, \Sigma\right)$ phase space is most easily visualized by plotting the level curves of ${\mathcal H}$ for different values of the constant parameter ${\mathcal L}$.\footnote{We have also confirmed that the same results are obtained through direct integration of the equations of motion.} It seems simplest to label the axes of the figures, using the  ``cartesian--type'' canonical variables:
\beq
U \;=\; \sqrt{2\left(\Sigma - {\mathcal L}\right)}\,\sin\Theta\,,\qquad\quad
V \;=\;  \sqrt{2\left(\Sigma - {\mathcal L}\right)}\,\cos\Theta\,,
\label{UVdef}
\eeq 

\noindent where $U$ and $V$ are new coordinates and momenta,
respectively. Before we discuss the phase space structure, it is
useful to interpret the canonical variables in terms of physical
quantities related to the centroids of the prograde and retrograde
populations:

\begin{itemize}

\item $\sqrt{U^2 + V^2} = \sqrt{2(\Sigma - {\mathcal L})} = \sqrt{2 {L}_{-}} = \sqrt{u_{-}^{2}+v_{-}^{2}}$ is proportional to the centroid eccentricity of the retrograde ring.

\item $U=V=0$ represents the zero eccentricity state; it is an equilibrium for ${\mathcal L}=0$, but not otherwise.

\item $\Theta = \left({\psi}_{+} \,+\, {\psi}_{-}\right) = \left(g^{\rm cent}_{-} \,-\, g^{\rm cent}_{+}\right)$ is the difference between the periapse angles of the centroids of the retrograde and prograde populations. Since we also have $\Theta = \arctan{(U/V)}$, we note that: 

\begin{itemize}
\item[(i)] $U=0$ and $V < 0$ implies that $\left(g^{\rm cent}_{-} \,-\, g^{\rm cent}_{+}\right) = \pi$, corresponding to eccentric $\pm$ discs with {\em anti--aligned} periapses.

\item[(ii)] $U=0$ and $V > 0$ implies that $\left(g^{\rm cent}_{-} \,-\, g^{\rm cent}_{+}\right) = 0$, corresponding to eccentric $\pm$ discs with {\em aligned} periapses.
\end{itemize}
\end{itemize}

\noindent
Before we begin, we need to determine the parameters --- other than ${\mathcal L}$ -- that control the dynamics. We first note that the constants $w_\pm$, $w_c$ and $\eta_\pm$ are all proportional to the total mass, $M$, in the $\pm$ discs. Each of the parameters $\left(\alpha, \beta, \gamma, \lambda, \kappa, \chi\right)$ is proportional to $a_0^{-3/2}\,$, where $a_0$ is the semi--major axis; these parameters are also functions of  the dimensionless softening parameter $r = b/a_0\,$, but the dependences are not so simple. Therefore, every term on the right side of equation~(\ref{hamsigthe}) is proportional to $Ma_0^{-3/2}\,$, so this combination of constants determines only the rate at which a phase trajectory is traversed. 
For the purposes of investigating the geometry of phase space, we may set
$Ma_0^{-3/2}$ equal to unity. The dimensionless masses, $\mu_+ = (M_+/M)$ and $\mu_- = (M_-/M)$, can both be expressed in terms of the dimensionless mass ratio, $\mu = (M_-/M_+)\,$. Therefore, we are left with the three dimensionless parameters $\left({\mathcal L}, r, \mu\right)$, which control the dynamics generated by ${\mathcal H}\,$. 

\subsubsection{Dynamics when ${\mathcal L} = 0$} 

When ${\mathcal L} = 0$, the prograde and retrograde discs have equal amounts of mass--weighted centroid eccentricities. We have already studied the linear instability of the zero eccentricity equilibrium in \S~\ref{sec:linear}. At a fixed value of $r$, there is a critical value of the mass ratio $\mu$ above which the instability sets it. When $r=0.1$, this critical mass ratio is ${\mu}_{\rm crit}\simeq 0.4436$, which corresponds to about $70\%$ of the disc mass in the prograde component and the remaining mass in the retrograde component. Here we explore the structure of nonlinear centroid dynamics as $\mu$ is varied, with both ${\mathcal L}$ and $r$ held fixed.

\begin{figure}
  \centering
  \begin{minipage}{1.0\linewidth}
    \centering
    \includegraphics[width=2.0in]{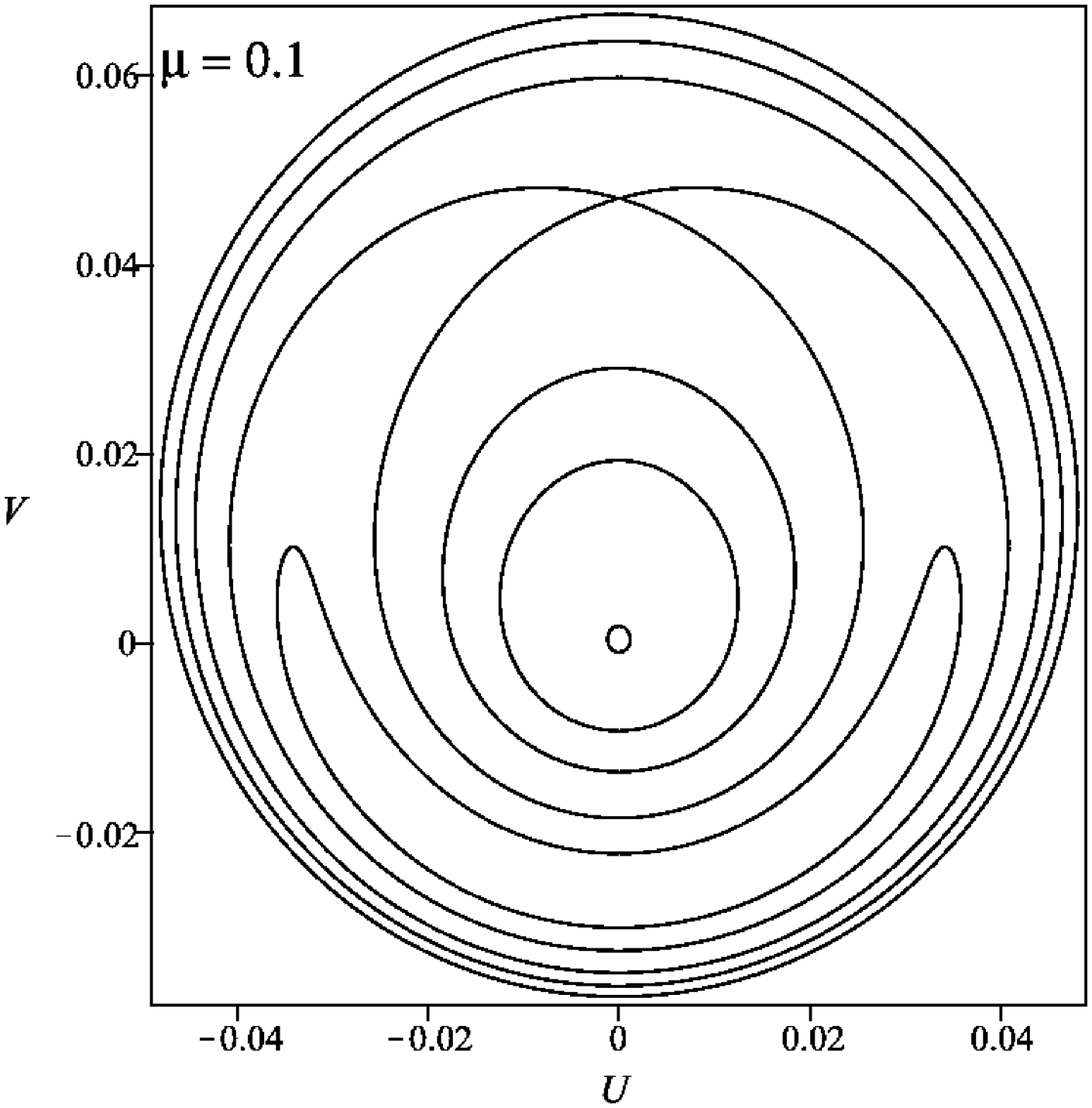}  
    \includegraphics[width=2.0in]{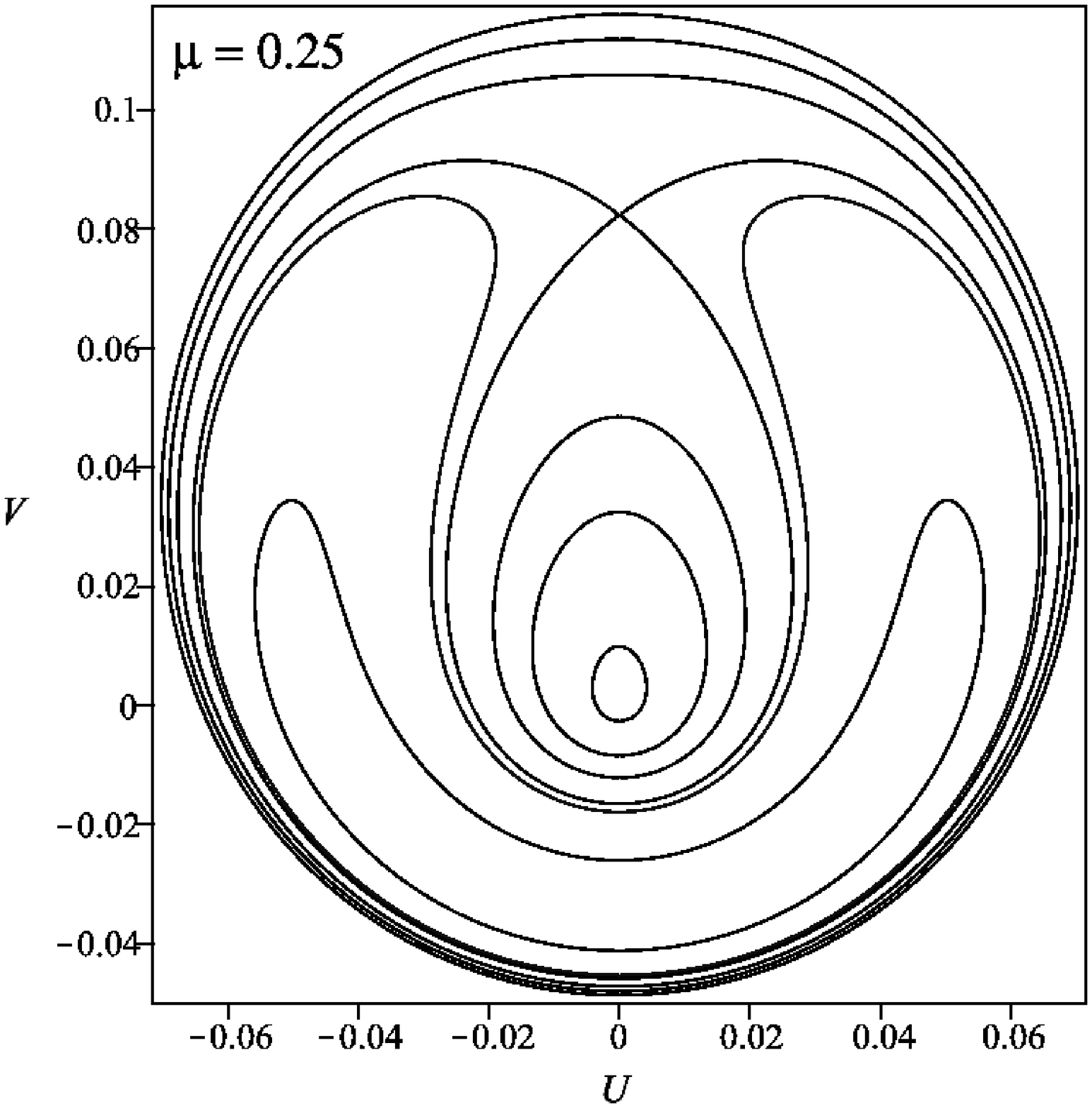}  
  \end{minipage}

\begin{minipage}{1.0\linewidth}
    \centering
    \includegraphics[width=2.0in]{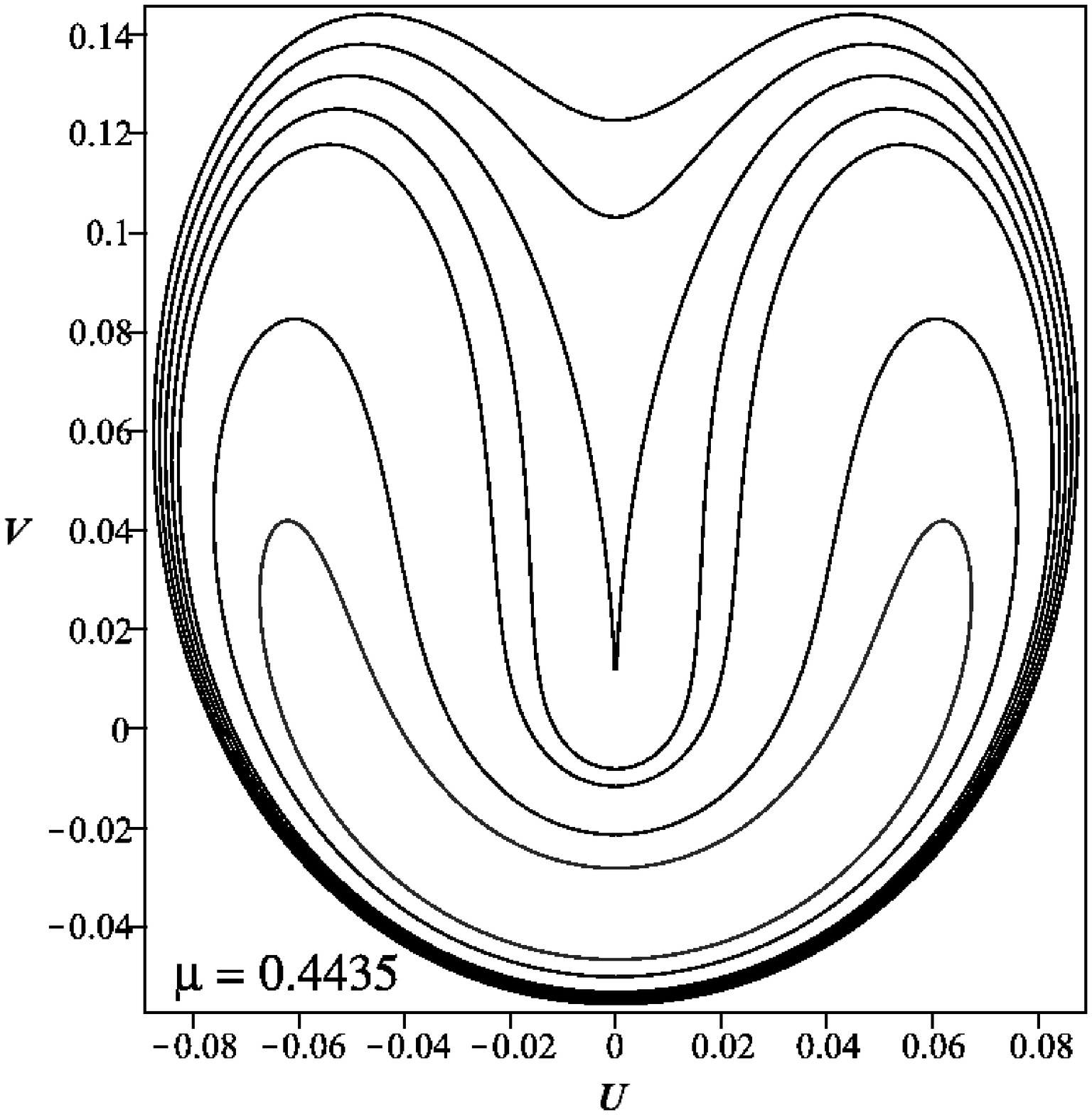}  
    \includegraphics[width=2.0in]{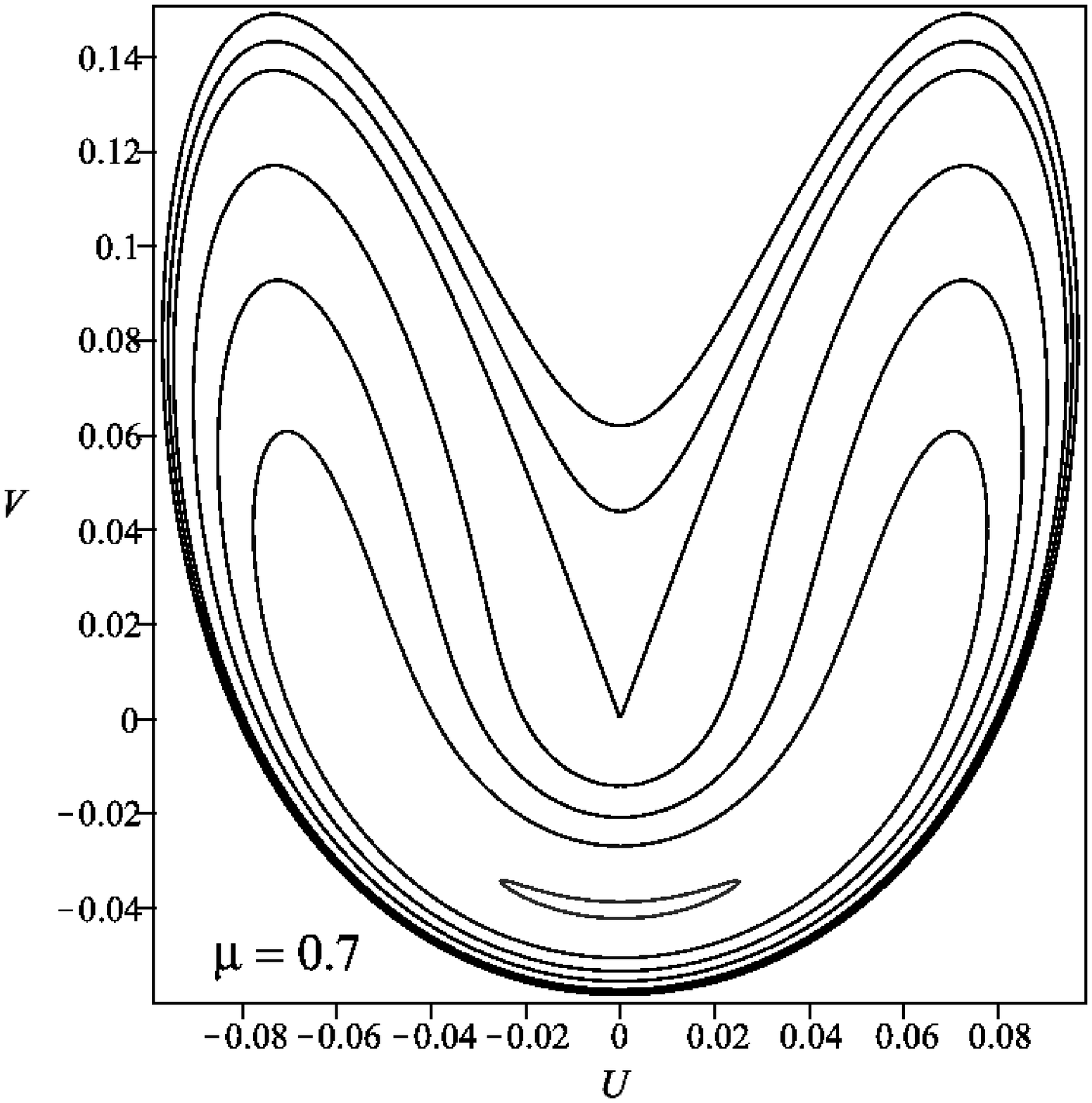}  
  \end{minipage}

\caption[Phase-Space]{Phase portraits for four different values of $\mu$, when ${\mathcal L} = 0\,$ and $r=0.1\,$.}
\label{fig:zero-ps}
\end{figure}

\begin{figure}
  \centering
  \begin{minipage}{1.0\linewidth}
    \centering
    \includegraphics[width=3.0in]{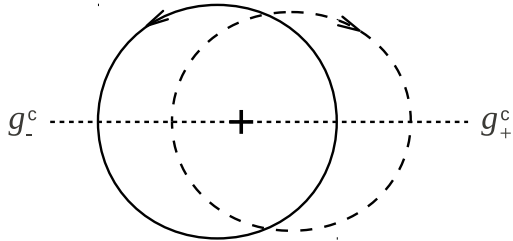}  
  \end{minipage}

\caption[ZeroL]{Snapshot of the $\pm$ centroid orbits in physical
space corresponding to the stable, precessing equilibrium P1 $(U=0,
V\simeq -0.04)$ in the bottom--right panel of
Fig.(\ref{fig:zero-ps}). The cross marks the location of the central
mass, which is at the common focus of both the prograde (solid) and the
retrograde (dashed) centroid orbits. The eccentricities of the $\pm$
orbits are $0.05$ and $0.06$, respectively, and have been greatly
exaggerated in the figures. Note that the $\pm$ periapses are
anti--aligned.}  \label{fig:centroids} \end{figure}

\begin{figure}
  \centering
  \begin{minipage}{1.0\linewidth}
    \centering
    \includegraphics[width=5in]{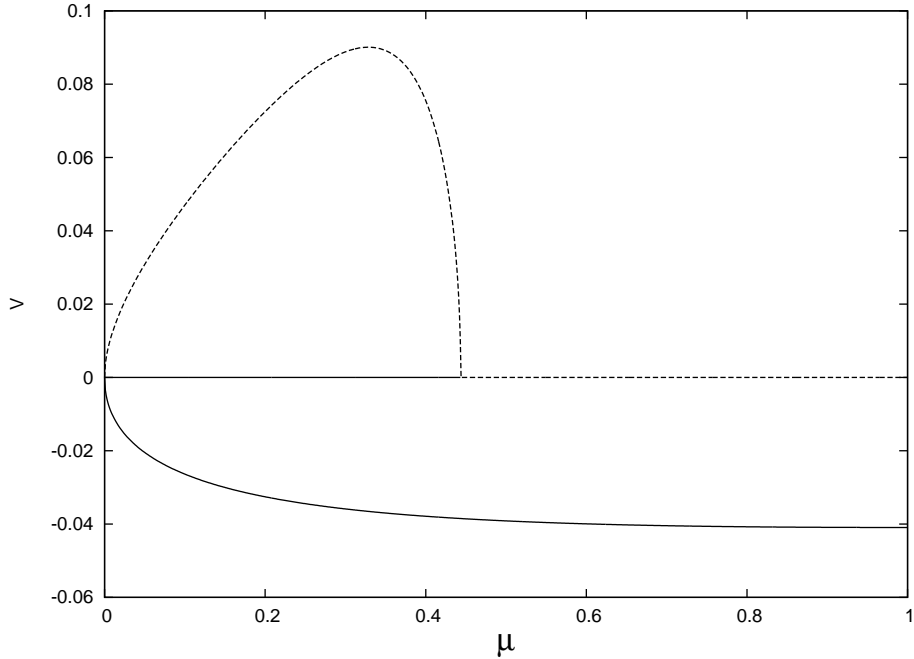}  
  \end{minipage}

\caption[Equilibria]{Bifurcation of equilibria: Plot of the values of the 
$V$--coordinates of the stable and unstable equilibria, as functions of the mass ratio $\mu$. Stable/unstable equilibria are shown in solid/dashed lines.}
\label{fig:zero-equi}
\end{figure}

\noindent
Fig.(\ref{fig:zero-ps}) displays the level curves of ${\mathcal H}$ in the $\left(U, V\right)$ phase space for four different values of $\mu$, when ${\mathcal L} = 0\,$ and $r=0.1\,$. Phase flows occur along these level curves. Some noteworthy features are: 

\begin{itemize}

\item For $\mu \leq {\mu}_{\rm crit}$, the zero--eccentricity equilibrium $P_2 = (0, 0)$ is stable; but there are also two additional equilibria, $P_1 = \left(0, V_1\right)$ and $P_3 = \left(0, V_3\right)$. 

\item $P_1$ is stable: it has $U=0\,$ and $V_1 < 0$, which corresponds to steadily precessing eccentric $\pm$ discs with anti--aligned periapses.  

\item $P_3$ is unstable: it has $U=0\,$ and $V_3 > 0$, which corresponds to steadily precessing eccentric $\pm$ discs with aligned periapses. 

\item When $\mu$ exceeds $\mu_{\rm crit}$, the zero--eccentricity equilibrium $P_2$ goes unstable by merging with the unstable equilibrium $P_3$. Small perturbations about $P_2$ now exhibit large variations in eccentricity, with phase space trajectories that take them on an excursion around the stable equilibrium $P_1\,$. A schematic view in physical space of anti-aligned centroid orbits at $P_1$ is shown in Fig.(\ref{fig:centroids}).

\item $P_1$ remains stable for all values of $\mu\,$.

\item The bifurcation of equilibria as a function of $\mu$, at fixed $r\,$, is shown in Fig.(\ref{fig:zero-equi}). The plot reflects transition across ${\mu}_{\rm crit}$, with the two equilibria $P_2$ and $P_3$ merging to form an unstable equilibrium, along with the continuing sequence of the stable equilibrium $P_1\,$.

\item Further increase in $\mu$ does not alter the equilibrium structure. However, there are larger excursions in eccentricity, so much so, that the conditions under which the model works can be in question.
\end{itemize}

\subsubsection{Dynamics when ${\mathcal L} \neq 0$} 

When ${\mathcal L} \neq 0$, the mass--weighted centroid eccentricities of the prograde and retrograde discs are unequal. Without loss of generality, we assume that ${\mathcal L} \geq 0$; in other words, we assume that the
mass--weighted centroid eccentricity of the prograde disc is greater than that of the retrograde disc. We follow the equilibria and their bifurcations as ${\mathcal L}$ is increased, at fixed $\mu$ and $r$. Some noteworthy features of the phase--space evolution, shown in Fig.(\ref{fig:nzero-ps}), are:

\begin{figure}
  \centering
  \begin{minipage}{1.0\linewidth}
    \centering
    \includegraphics[width=2.0in]{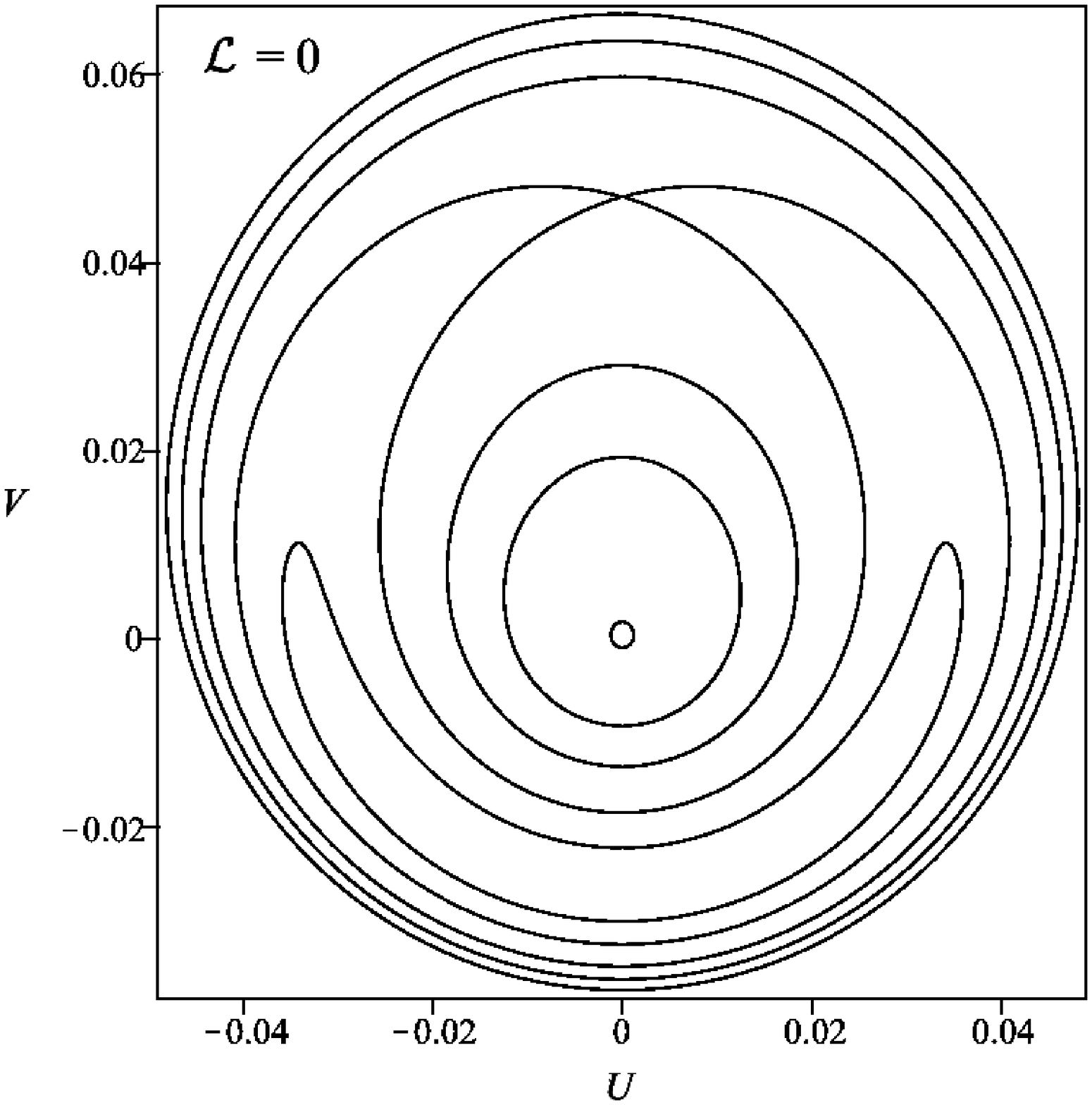}  
    \includegraphics[width=2.0in]{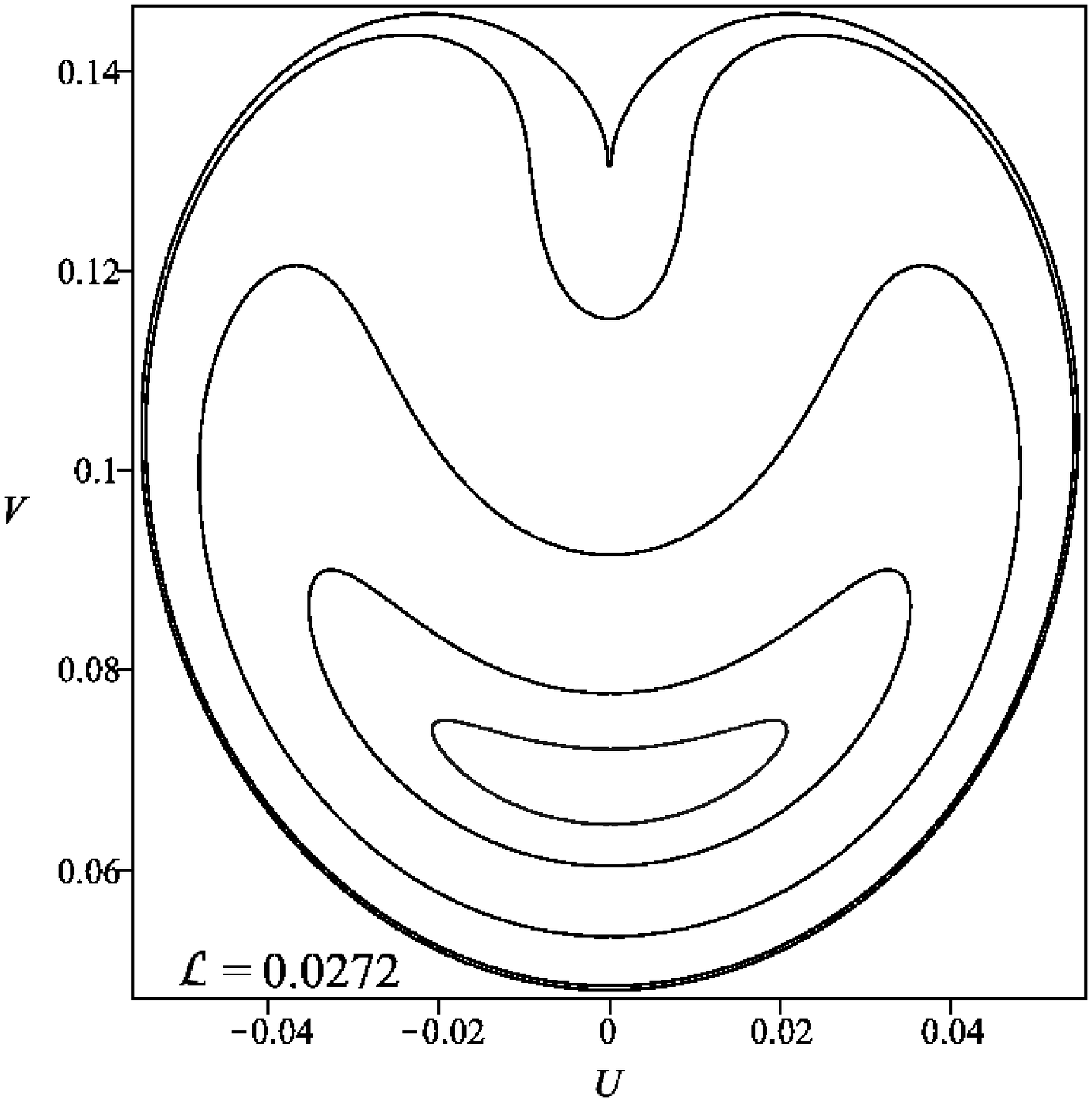}  
  \end{minipage}

  \begin{minipage}{1.0\linewidth}
   \centering
   \includegraphics[width=2.0in]{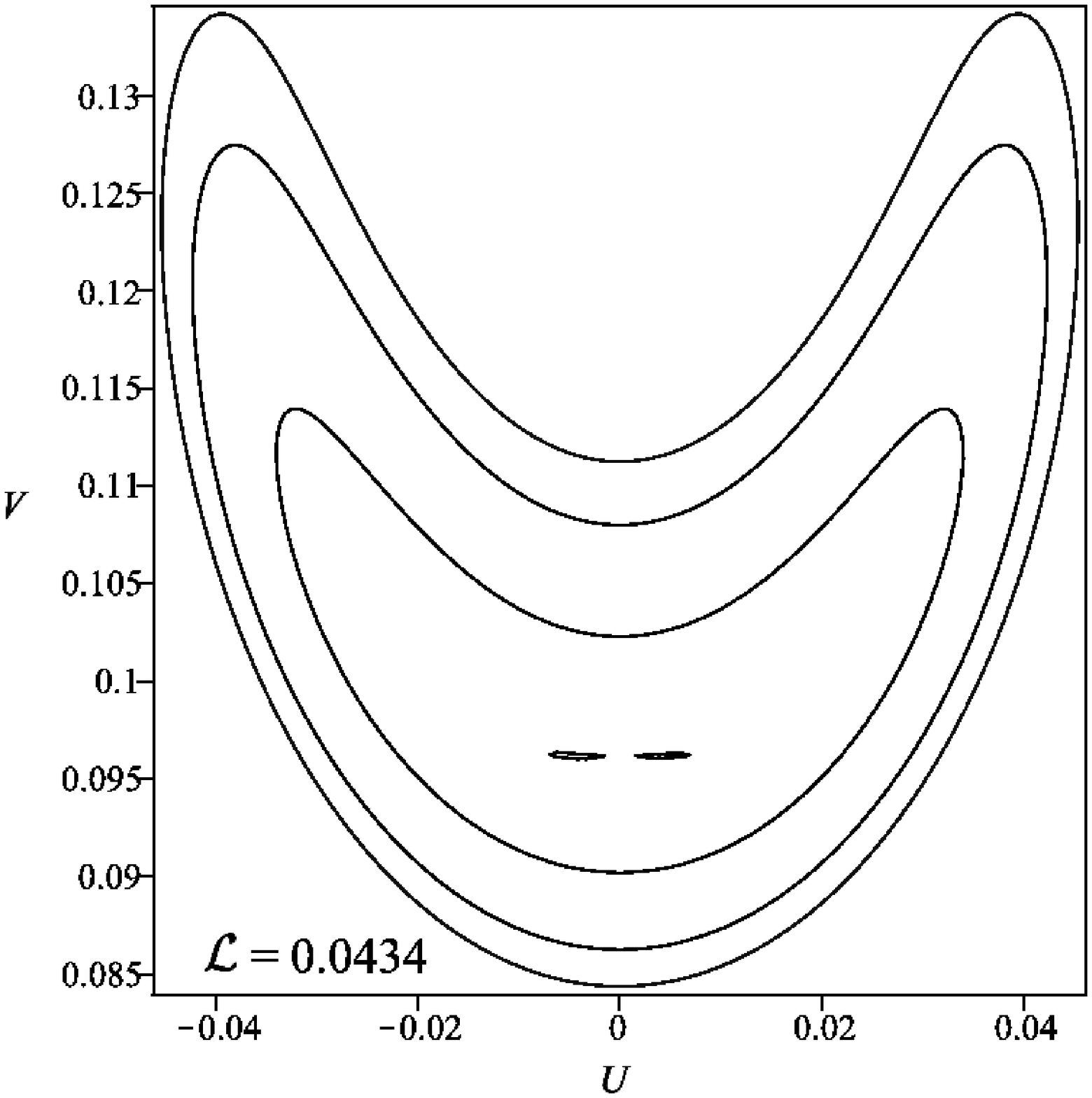}  
   \includegraphics[width=2.0in]{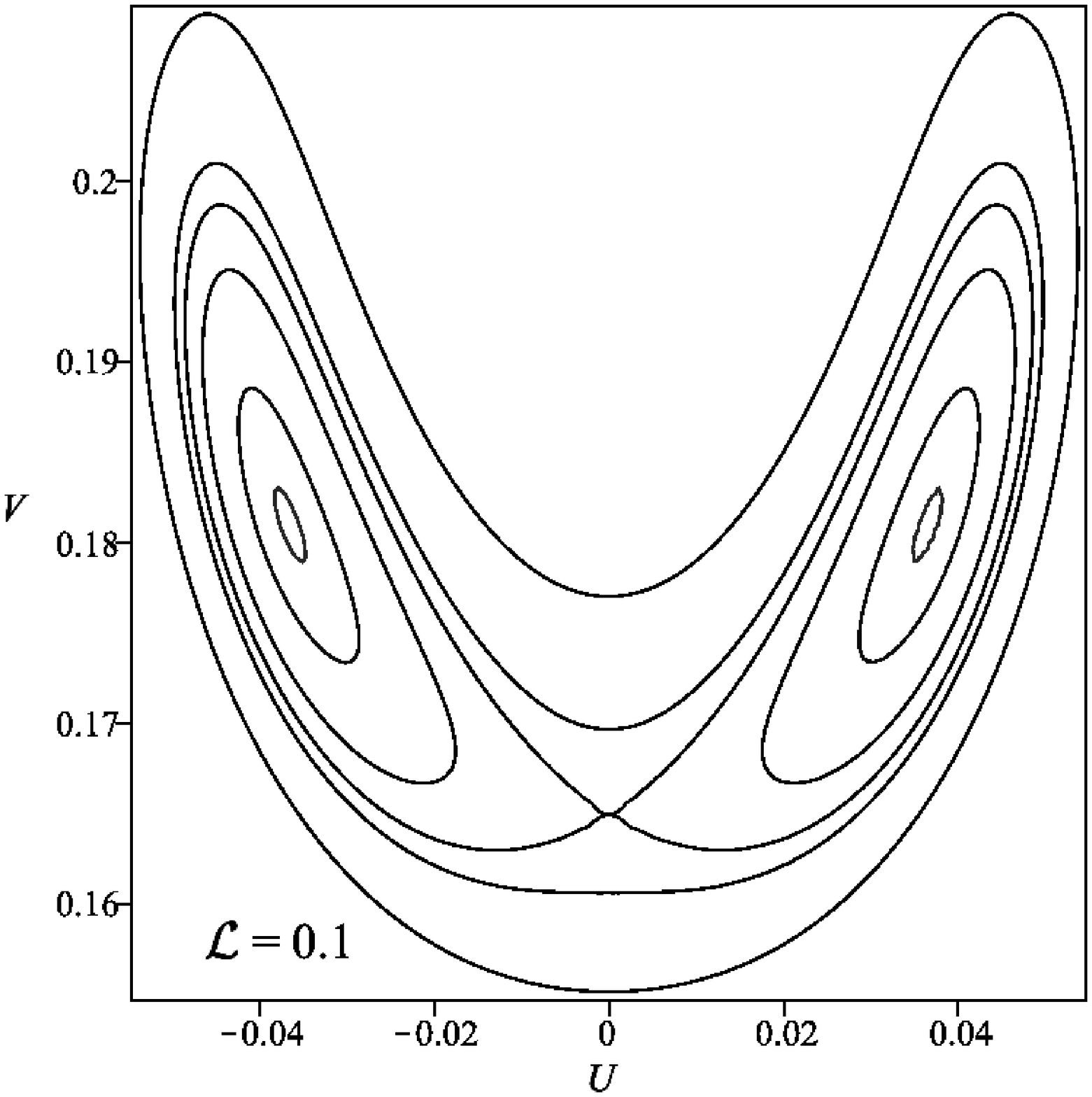}  
   \end{minipage}
\caption{Phase--space evolution with increasing value of ${\mathcal L}$.}
\label{fig:nzero-ps}
\end{figure}

\begin{figure}
  \centering
  \begin{minipage}{1.0\linewidth}
    \centering
    \includegraphics[width=3.0in]{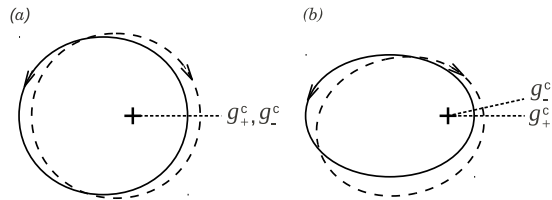}  
  \end{minipage}
\caption[NonZeroL]{Snapshots of the $\pm$ centroid orbits in physical space, corresponding to the stable (precessing) equilibrium P1 for two different cases. The cross marks the location of the central
mass, which is at the common focus of both the prograde (solid) and the retrograde (dashed) centroid orbits. Panel "a" corresponds to top--right panel of Fig.(\ref{fig:nzero-ps}) ($\cal{L}$= 0.0272), when P1 $(U=0, V\simeq 0.07)$ is still a stable equilibrium before bifurcation. The eccentricities of the $\pm$ orbits are $0.35$ and $0.2$, respectively. Note that the $\pm$ periapses are aligned. Panel "b" corresponds to the bottom--right panel of Figure~4 ($\cal{L}$= 0.0272), when P1 has bifurcated into two stable equilibria, P1a and P1b, which are at $U\simeq \pm 0.04$ and  $V\simeq 0.18$; we show orbits for P1a . The eccentricities of the $\pm$ orbits are $0.65$ and $0.55$, respectively. Note that the $\pm$ periapses are mis-aligned, with the pericentre of the ``-'' orbit leading the pericentre of the ``+'' orbit by about $12.5\deg$ (for P1b, the  pericentre of the ``-'' orbit would lag the pericentre of the ``+'' orbit by the same amount.}
\label{fig:nonzeroL}
\end{figure}

\begin{figure}
  \centering
  \begin{minipage}{1.0\linewidth}
    \centering
    \includegraphics[width=3.0in]{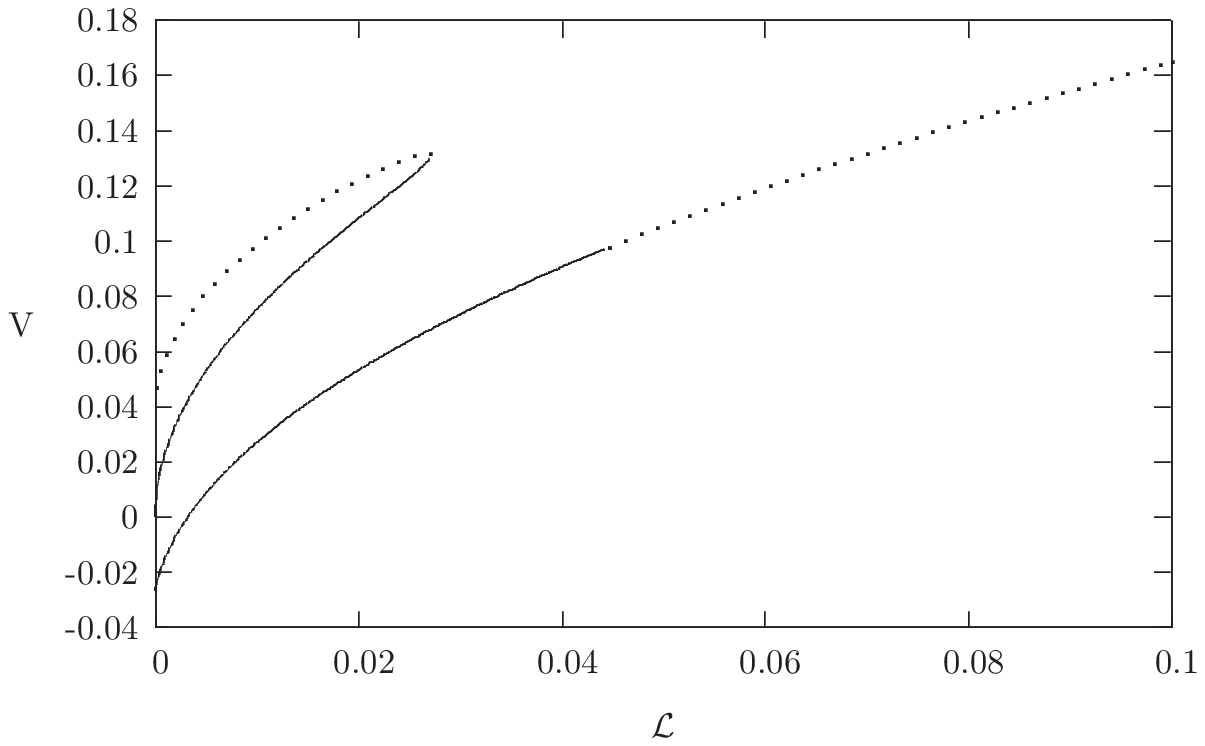}  
    \includegraphics[width=3.0in]{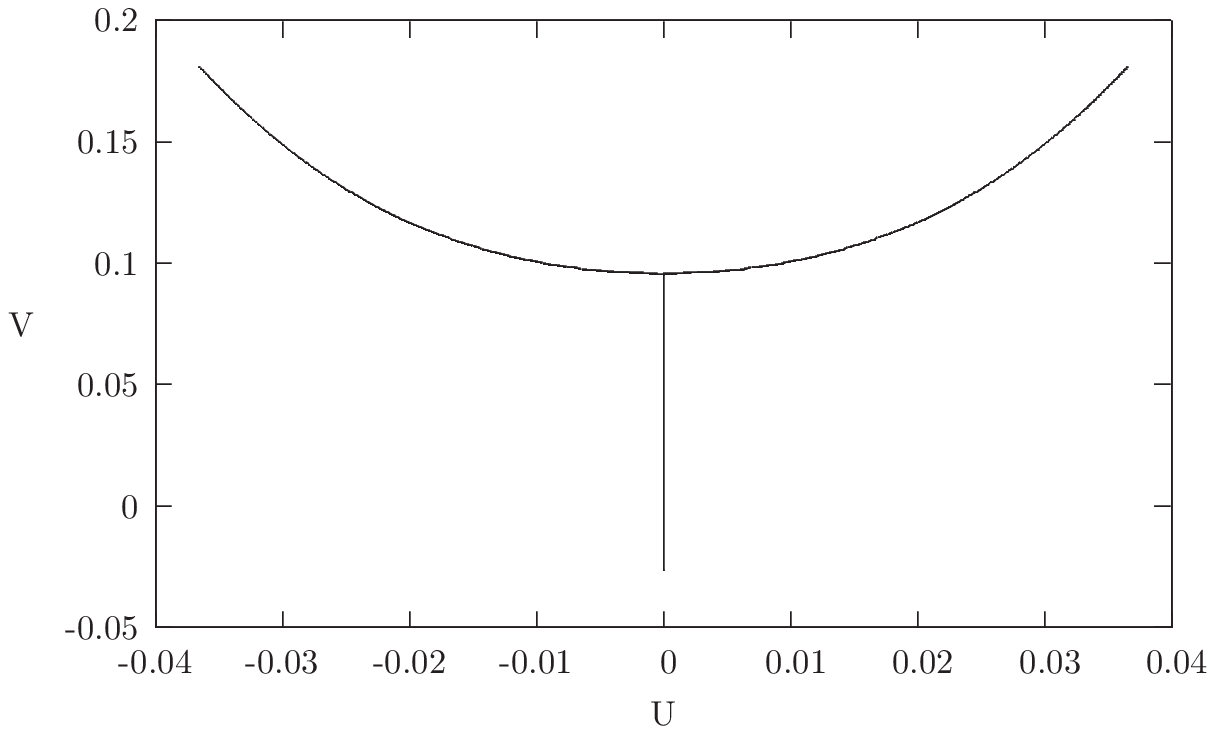}  
  \end{minipage}

\caption[Equilibria]{Uniformly precessing configurations for ${\mathcal L} \neq 0\,$. {\em Left}: Plot of the $Y$--coordinates of the equilibria as
functions of ${\mathcal L}$, for the on--axis ($U=0$) equilibria which correspond to aligned and anti--aligned periapses. Stable/unstable equilibria are shown in solid/dashed lines, unstable. Two critical values of ${\mathcal L}$ are apparent: ${\mathcal L} \simeq 0.0271$, at which stable and unstable equilibria merge; ${\mathcal L} \simeq 0.0433$, at which $P_1$ becomes unstable, giving birth to non--aligned equilibria (with $U \neq 0\,$).
Also apparent is the transition, at ${\mathcal L} \simeq 0.0031$, of the $P_1$, from being an equilibrium with stable anti--aligned periapses ($V_1 < 0\,$) to stable aligned periapses ($V_1 > 0\,$). {\em Right:} The bifurcation of the $P_1$, from being an aligned on--axis equilibrium to non--aligned equilibria.}
\label{fig:nzero-bif}
\end{figure}

\begin{itemize}

\item As we have seen earlier, when ${\mathcal L} = 0\,$, there is a qualitative change in the phase portrait when $\mu$ is smaller or larger
than $\mu_{\rm crit}\,$, for some chosen value of $r$. For $r=0.1$, $\mu = 0.1 < \mu_{\rm crit}\,$ so the zero--eccentricity equilibrium: we have three equilibria; two stable ($P_1$, $P_2$) and one unstable ($P_3$), as
discussed above.

\item With increasing ${\mathcal L}$, $P_2$  (which was initially at the origin) shifts continuously to higher eccentricity with $U_2 = 0$ and $V_2 > 0\,$, corresponding to steadily precessing eccentric $\pm$ discs with aligned periapses. At the critical value of ${\mathcal L} \simeq 0.0271\,$, $P_2$ and $P_3$ collide. 

\item All through, $P_1$ remains stable with $U_1 = 0\,$. However, as ${\mathcal L}$ increases, $V_1$ also increases and, near ${\mathcal L} \simeq 0.0031\,$, $V_1$  becomes positive from its initially negative value; thus the corresponding stable, steadily precessing eccentric $\pm$ discs switch from anti--aligned to aligned periapses. 

\item With further increase in ${\mathcal L}$, $P_1$ remains stable with $U_1 = 0$ and increasing $V_1$ until we hit another critical value, ${\mathcal L} \simeq 0.0433$\,. At this value of ${\mathcal L}$, $P_1$ becomes unstable and, in a pitchfork--like bifurcation, there emerge two stable and {\em non--aligned} equilibria. The remarkable feature of these new equilibria is that they are neither aligned nor anti--aligned. This suggests that, for large enough values of ${\mathcal L}$, the stable, uniformly precessing counter--rotating discs have periapses that are neither aligned nor anti--aligned. The transition from aligned to non-aligned stable equilibria at the bifurcation is illustrated in Fig.(\ref{fig:nonzeroL}) with centroid orbits in physical space, at $P_1$ before the bifurcation, and at one of its two stable offsprings after.

\item Increasing ${\mathcal L}$ still further maintains the equilibrium structure with the stable equilibria increasing in eccentricity and  misalignment, in a fashion reminiscent of the displacement of an unstable bead on a rapidly spinning hoop. 

\item  In the left panel of Fig.(\ref{fig:nzero-bif}), we show the on--axis equilibria, the eventual merging of $P_2$ and $P_3$, followed by the loss of stability of $P_1$. In the right panel, we follow the bifurcation from $P_1$ (now unstable) of the two stable, non--aligned equilibria.
\end{itemize}

\section{Summary and conclusions}

We have formulated the problem of the collisionless, self--consistent
dynamics of nearly Keplerian star clusters around a massive black
hole. Averaging over the fast Keplerian phase, we used the result that
the semi--major axes of the stellar orbits are nearly conserved
quantities. Hence each stellar orbit may be imagined to be a
\emph{Gaussian ring}, of fixed semi--major axis, whose shape and
orientation changes over time scales that are longer than the Kepler
orbital times. Since the semi--major axis of each stellar orbit is
constant, and the orbital phase unimportant, the phase space in which
the \emph{slow} dynamics of an individual orbit occurs is
four--dimensional. In terms of the Delaunay actions--angle variables,
the two actions are the magnitude of the angular momentum and the
$z$--component of the angular momentum, with corresponding angles
being the argument of the periapse and the longitude of the ascending
node. The star cluster was described by a distribution function (DF),
which is a function of five variables: the four action--angle
variables described above, and the constant semi--major axis which can
differ from star to star. We presented the collisionless Boltzmann
equation (CBE), which governs the self--consistent slow dynamics of
the star cluster. The Hamiltonian determining this dynamics is the
orbit--averaged gravitational potential energy between the stars; it
is determined by integrating a softened Keplerian potential over the
orbital phases of both rings (thereby forming a ``ring--ring
interaction function''), and then integrating this over all of phase
space, suitably weighted by the DF.

The goal of this paper is to explore the counter--rotating
instability. To this end, we considered the CBE for razor--thin,
planar discs; in this case, at each value of the semi--major axis, the
phase space is two dimensional, topologically equivalent to the
2--sphere. Then we considered the counter--rotating discs as two
separate collisionless populations of stars, the prograde (or ``$+$'')
disc and the retrograde (or ``$-$'' disc). For convenience, we assumed
that all the prograde stars have the same semi--major axes and all the
retrograde stars have the same semi--major axes. Thus the total phase
space is the direct sum of two 2--spheres; one for the prograde disc,
the other for the retrograde disc. We then wrote down two separate
$\pm$ DFs which obeyed two separate $\pm$ CBEs, governed by two
separate $\pm$ Hamiltonians. Since the $\pm$ populations are
gravitationally coupled, each of the $\pm$ Hamiltonians depends on
both the $\pm$ DFs. When the discs are composed of stellar orbits of
small eccentricities, the prograde population is clustered around the
north pole of its 2-spherical phase space, and the retrograde
population is clustered around the south pole of its 2-spherical phase
space. We transformed to new ``cartesian--type'' canonical variables,
which are more convenient to use in the limit of small
eccentricities. The ring--ring interaction function was then expanded
to fourth order in the eccentricities, using results from
\cite{mt11}. So, each of the $\pm$ Hamiltonians was obtained as a
quartic polynomial of the $\pm$ canonical coordinates and momenta,
with coefficients that depend on both the $\pm$ DFs. 

Time--dependent DFs were constructed using Jeans' theorem. The first
step was to identify appropriate phase space functions which are
approximately conserved by the time--dependent $\pm$
Hamiltonians. Canonical transformations to moving origins in the $\pm$
phase spaces were used to construct these approximate dynamical
invariants; to lowest order in the eccentricities, these are quadratic
functions of the phase space coordinates and momenta with time
dependent coefficients. The next step was to choose the $\pm$ DFs to
be some physically allowed function (positive, finite mass, low
eccentricity stars etc.) of the $\pm$ invariants. The DFs are such
that their isocontours in the $\pm$ phase spaces are ellipses,
centered on moving origins. The time--dependent coordinates of the
origins were referred to as the \emph{centroids}. The evolving shapes
and orientations of the ellipses were described by $\pm$ \emph{shape
matrices}, which are time--dependent and positive--definite; these
matrices describe the anisotropic nature of the dispersion in the
eccentricities. It is important to note that the DFs so constructed
are approximate time--dependent solutions, when the dispersions of
eccentricities described by the shape matrices are less than the
centroid eccentricities. Thus the coupled, time--dependent dynamics of
the the counter--rotating discs is described by a set of ordinary
differential equations (ODEs) describing the evolution of the
centroids and shape matrices. Some general properties are:

\begin{enumerate}

\item For each of the $\pm$ populations, there are two centroid coordinates 
and one positive--definite $2\times 2$ shape matrix. Thus we have a system of
12 first order ODEs describing counter--rotating disc dynamics. 

\item The centroid equations are a set of 4 autonomous (i.e. time--independent) 
first order ODEs. These are independent of the shape matrices, because we are
working in the limit where the centroid eccentricities are greater than the
dispersion of eccentricities. 

\item The centroid equations conserve the total angular momentum corresponding
to the centroid eccentricities.

\item The equations for the shape matrices are linear ODEs which depend 
on the centroid coordinates. Thus shape dynamics is driven by centroid dynamics.

\item The shape equations conserve the determinants of the shape matrices, so we may choose the determinants to be equal to unity.

\item The 4 autonomous first order ODEs describing centroid dynamics have 
cubic nonlinearity. However, quite remarkably, they constitute an integrable system. This happens because the 4--dimensional system corresponds to a
2--degree of freedom Hamiltonian system which admits two conserved quantities; the centroid angular momentum and the Hamiltonian itself.

\end{enumerate}

We then studied the linear stability of initially zero--eccentricity discs, 
and derived the conditions under which the configuration is unstable. Some notable properties are:

\begin{enumerate}  

\item For a given softening, there is a critical value of the mass ratio (which decreases with increasing softening) above which the zero--eccentricity equilibrium is unstable to the growth of eccentricities in both $\pm$ discs. 

\item Conversely, for a given mass ratio, there is a critical value of the softening (which increases with decreasing mass ratio) above which the zero--eccentricity equilibrium is unstable. 

\item The stable solutions correspond to normal modes describing steadily precessing discs of fixed centroid eccentricities. 

\item When the parameters are in the unstable regime, both growing and damped solutions are allowed. The growing (damped) solution describes discs whose eccentricities grow (damp) as they precess steadily.

\item The physical basis of the instability is through exchange of angular momentum between the $\pm$ discs. When the prograde disc gives some angular momentum to the retrograde disc, both discs increase their eccentricities. For the instability to operate, the mass ratio has to be large enough to be able to overcome the effective ``heat'' in the stellar distribution which, in our cold DFs, is mimicked by the softening.

\end{enumerate}

The nonlinear dynamics is, of course, much richer. We demonstrated that the
ODEs of centroid dynamics could be cast into Hamiltonian form, with a 2--degree
of freedom, time--independent Hamiltonian which is quartic in the canonical
variables. This Hamiltonian conserves two independent phase space functions; the total angular momentum of the centroid dynamics and the Hamiltonian itself.
Therefore, the nonlinear dynamics of centroid motion is completely integrable.  
We exploited the conservation of the angular momentum to reduce the dynamics
to that of a 1--degree of freedom system, where the angular momentum appears
as a constant parameter; specifically, we used the quantity ${\mathcal L}$, which is the amount by which the angular momentum is lower than the maximum
value which is attained for discs with zero centroid eccentricities. The 
results from a preliminary exploration of the nonlinear dynamics are given below.

\noindent
I. Case ${\mathcal L}=0$:

\begin{enumerate}

\item A special case includes the initially zero--eccentricity discs, 
whose linear instability was discussed earlier. We followed the global phase
space structure, as a function of varying mass ratio, $\mu$, at fixed softening. When $\mu < \mu_{\rm crit}$, the zero eccentricity state is, of course, stable. 

\item In addition, there are two other equilibria, one stable and the other unstable, both of which correspond to eccentric $\pm$ discs which precesses steadily. The unstable (stable) equilibrium corresponds to the discs having aligned (anti--aligned) periapses. The stable equilibrium remains stable for all values of $\mu\,$.

\item When $\mu$ exceeds $\mu_{\rm crit}$, the zero--eccentricity equilibrium goes unstable by merging with the unstable equilibrium.
\end{enumerate}

\noindent
II. Case ${\mathcal L} \neq 0$:

\begin{enumerate}

\item We followed the global phase space structure, as a function of varying 
${\mathcal L}$, at fixed mass ratio and softening. These fixed parameters were
chosen such that the zero--eccentricity discs would have been in the stable regime; specifically, we chose $\mu = r = 0.1\,$.

\item When ${\mathcal L}=0$, there are three equilibria, two stable and one 
unstable, as discussed above.

\item With increasing ${\mathcal L}$, the zero--eccentricity equilibrium
remains an equilibrium point, but now corresponds to steadily precessing eccentric $\pm$ discs with aligned periapses. At the critical value of ${\mathcal L} \simeq 0.0271\,$, it merges with the unstable equilibrium. 

\item Meanwhile, the other stable equilibrium remains stable as ${\mathcal L}$ increases. However, near ${\mathcal L} \simeq 0.0031\,$, the steadily
precessing eccentric $\pm$ discs switch from anti--aligned to aligned periapses. Then, near ${\mathcal L} \simeq 0.0433$, the equilibrium becomes unstable and, in a pitchfork--like bifurcation, there emerge two stable and {\em non--aligned} equilibria. The remarkable feature of these new equilibria is that they are neither aligned nor anti--aligned. This suggests that, for large enough values of ${\mathcal L}$, the stable, uniformly precessing counter--rotating discs have periapses that are neither aligned or anti--aligned. 

\item Increasing ${\mathcal L}$ still further maintains the equilibrium structure with the stable equilibria increasing in eccentricity and  misalignment, in a fashion reminiscent of the displacement of an unstable bead on a rapidly spinning hoop. 
\end{enumerate}

The brief report above is but a preliminary account of the vast and
rich dynamics of this gravitationally coupled system. Here, we have
attempted a self--contained presentation, whose aim is to point out
the variety of steadily precessing eccentric configurations that are
allowed and how their properties and stability depend on parameters
such as the disc mass ratios and angular momentum. Straightforward
generalizations are possible for discs with different values of the
semi--major axes, or possibly a range of values of the semi--major
axes. Razor thin ${\pm}$-disks, with a spread in semi-major axes, are
already known to exhibit some of the dynamical properites of the
singular distributions studied here: a- a linear stability threshold
which reflects an interplay between softening (heat) and ${\pm}$-mass
ratio (self-gravity) \citep{tou02, ss10}; the possibility of a
non-aligned, uniformly precessing equilibrium, first identified by
\citet{ss02} as a promising stellar dynamical model of the double
nucleus of M31. A multi-centroid generalization of our two-centroid
theory, one which allows for a range in the semi-major axes, is
expected to provide idependent confirmation of these results, and
might, in addition, exhibit new and unsuspected dependences on the
semi--major axes. Other generalizations involve the (combined) effects
of a central density cusp and general relativistic corrections, both
of which induce apse-precession that can be strong enough to alter the
stability threshhold dramatically. A limitation of the work in this
paper is that we have not been able to discuss questions regarding the
saturation of instabilities, such as those explored in
\citet{ttk09}. To attempt such a description would involve taking any
of the time--dependent DFs of this paper as an ``unperturbed''
solution, and solving the linearized CBE for perturbations about this
state. Such a saturated state could well describe the
double--peaked, lopsided nuclear disc of M31.

Whereas we have focused attention on counter--rotating nearly
Keplerian discs, our general formulation of the CBE for nearly
Keplerian star clusters extends the work of \citet{st99} to
self--gravitating systems. In particular, our formalism of \S~2
applies to fully three dimensional clusters. In addition to the
eccentricity--periapse dynamics which applies to planar discs, we will
also have to consider the inclination--node degrees of freedom. An
expansion of the Hamiltonian to fourth order in the (sine of the)
inclinations will need only a modest extension of the methods
introduced in this paper. Then, it will be possible to explore the
stellar dynamics of counter--rotation, eccentricity and inclination
all considered together for nearly Keplerian systems.

\section{Acknowledgments} We thank the referee James Binney for a
thoughtful report which helped improve the quality of the
paper. J.R.T. acknowledges stimulating discussions with Abdul Hussein
Mrou\'eh. We thank Nishant K. Singh for assistance with some figures.

\appendix
\section{Ring--ring interaction function to $4^{\rm th}$~order in the eccentricities}

We provide explicit expressions for the expansion of the orbit
averaged interaction function between two softened, coplanar Gaussian
rings, given in equation~(\ref{phifull}), up to 4th order in the
eccentricities. The expansion is carried out for arbitrary semi-major
axes with the help of classical techniques of celestial mechanics,
generalized to softened interactions. These same methods can be used
to recover expansions to arbitrary orders in eccentricity and
inclination. Details of the techniques, expansions, accuracy and
conditions for convergence are discussed in \citet{mt11}.

We consider two coplanar rings with orbital elements $(a, e, g)$ and $(a',e',g')$, define 

\beq 
\rho \;=\; {\rm min} (a,a')/{\rm max} (a,a')\,; \qquad
r \;=\; b/ {\rm max} (a,a')\,, 
\eeq

\noindent 
and denote the softened analog of the classical Laplace coefficients by,

\begin{equation}
B^m_s(\rho, r)  =\frac{2}{\pi} \int^\pi_0 \frac{\cos (m\, t) \,dt}
{ {\left[1 + \rho^2 + r^2  - 2 \rho  \cos (t) \right]}^{s/2}}.
\end{equation}

\noindent The 4th order expansion (with the constant ignored) takes
the form:

\begin{eqnarray}
\Psi & = & \frac{\rho}{2{\rm max} (a,a')} \Big\{c_{20}^{0} e^2  +   c_{11}^{1}  ee' \cos (g - g')  + c_{02}^{0} e'^2 + c_{40}^{0} e^4  +  c_{31}^{1} e^3 e'\cos (g - g') \nonumber \\[1em]
&  &  + c_{22}^{0} e^2 e'^2 + c_{22}^{2} e^2 e'^2 \cos [2 (g - g')]  +   c_{13}^{1} e e'^3\cos (g - g') + c_{04}^{0} e'^4 \Big\},  
\label{psi-4th}
\end{eqnarray}

\noindent
where

\begin{eqnarray}
  c_{20}^{0} &=& - \Big[ \frac{3}{4} r^2 ( \rho B_5^0 - B_5^1) - f(\rho,r) \Big]\,, \nonumber \\[1em]
  c_{11}^{1} &=& - \Big[  - \frac{9}{4} B_3^0 - \frac{1}{4} B_3^2 + \frac{9}{4} ( 1 + \rho^2 ) B_5^0 
        - \frac{21}{8} \rho B_5^1 - \frac{3}{4}  ( 1 + \rho^2 ) B_5^2 - \frac{3}{8} \rho B_5^3  \Big] \,, \nonumber \\[1em]
  c_{02}^{0} &=& - \Big[ \frac{3}{4} \frac{r^2}{\rho} (B_5^0 - \rho B_5^1) - f(\rho,r) \Big]\,,
\end{eqnarray}

\noindent
with
 
\begin{equation}
f(\rho,r) = -\frac{5}{4} B_3^1 + \frac{3}{8}\rho B_5^0 + \frac{3}{4} (1+\rho^2)B_5^1 - \frac{15}{8}\rho B_5^2 \,,
\end{equation}

\noindent
and
 
\begin{eqnarray}
  c_{40}^{0} &=&
   \frac{1}{192} 
  \Big[  -3 B_3^1  +  \frac{9}{2} \rho (1 + 30 \rho^2) B_5^0 
      + 180 \rho^2  B_5^1  + \frac{423}{2} \rho B_5^2 - 225 \rho^3 ( 5 +2 \rho^2) B_7^0 \nonumber \\[1em] 
        && \qquad + 45 \rho^2 ( -17 + 16 \rho^2) B_7^1 + 1035 \rho^3  B_7^2 + 585 \rho^2 B_7^3 
      + \frac{315}{8} \rho^3 (59 + 56 \rho^2 +8 \rho^4) B_9^0  \nonumber 
      \\[1em] 
        && \qquad - 315 \rho^4 ( 7 +4 \rho^2) B_9^1  - \frac{315}{2} \rho^3 (15 +2 \rho^2) B_9^2
      + 945 \rho^4 B_9^3 + \frac{2835}{8} \rho^3 B_9^4 \Big] \,, 
\end{eqnarray}

\begin{eqnarray}
  c_{31}^{1} &=&  \frac{1}{16}  
  \Big[ -3 B_3^2 - \frac{135}{2} \rho^2 B_5^0 
    -  60 \rho B_5^1 
    - \frac{3}{2} ( 6 + 11 \rho^2 ) B_5^2 
    - 24 \rho B_5^3 
    + \frac{15}{4} \rho^2 ( 109  + 60 \rho^2) B_7^0  \nonumber \\[1em]
    & &   \qquad 
    - \frac{15}{4} \rho ( -21 +68 \rho^2 )  B_7^1  
    - \frac{45}{2} \rho^2 ( 11 + 2 \rho^2) B_7^2  
    - \frac{15}{4} \rho ( 13 + 20 \rho^2 ) B_7^3 
    - \frac{165}{4} \rho^2 B_7^4 
      \nonumber \\[1em]  & &  \qquad 
    - \frac{105}{8} \rho^2 ( 33 + 59 \rho^2 + 12 \rho^4) B_9^0  
    + \frac{105}{8} \rho^3 (43 +34 \rho^2) B_9^1 + \frac{105}{2} \rho^2 ( 7 +4 \rho^2 +  \rho^4 ) B_9^2
        \nonumber \\[1em]  
        & &  \qquad  
    - \frac{105}{16} \rho^3 ( 19 + 4 \rho^2 ) B_9^3 
    - \frac{105}{8} \rho^2 ( 3 +5 \rho^2) B_9^4 
    - \frac{315}{16} \rho^3 B_9^5 \Big] \, ,
\end{eqnarray}

\begin{eqnarray}
  c_{22}^{0} &=& \frac{1}{128}
  \Big[ 32 B_3^1 + 384 \rho B_5^0 
    + 240 ( 1+ \rho^2 ) B_5^1
    + 552\rho B_5^2 
    - 2220 \rho ( 1+ \rho^2 ) B_7^0  \nonumber \\[1em]
    & & \qquad 
        - 240  ( -1 +\rho)^2 (1+\rho)^2 B_7^1
      + 1500 \rho ( 1 + \rho^2 ) B_7^2 + 1440 \rho^2 B_7^3 
      \nonumber \\[1em]  
& &   \qquad 
      + 105 \rho ( 20 +83 \rho^2 + 20 \rho^4 ) B_9^0  - 4620 \rho^2 (1+\rho^2) B_9^1  
    \nonumber \\[1em]
& &   \qquad 
    - 420 \rho ( 3 +11 \rho^2 +3 \rho^4) B_9^2  + 1260 \rho^2 ( 1 + \rho^2 )  B_9^3 
    + 945 \rho^3 B_9^4 \Big] \,, 
\end{eqnarray}

\begin{eqnarray}
  c_{22}^{2} &=&  -  \frac{1}{512}
  \Big[ -8 B_3^1 - 72 B_3^3 - 2388 \rho B_5^0 
    + 120 ( 1 + \rho^2 ) B_5^1 + 24 \rho B_5^2 
    - 216 ( 1+ \rho^2) B_5^3    \nonumber\\[1em]  
        & & \qquad 
       - 276 \rho B_5^4 + 6300 \rho ( 1 + \rho^2) B_7^0 
       - 120 (1+75\rho^2 + \rho^4) B_7^1
       - 360 \rho ( 1 + \rho^2) B_7^2 \nonumber \\[1em]  
        & & \qquad 
       - 60 ( 6 + 5 \rho^2 + 6 \rho^4 ) B_7^3 - 660 \rho ( 1 + \rho^2 ) B_7^4   
       - 300 \rho^2 B_7^5 - 3780 \rho ( 1+ \rho^2 + \rho^4) B_9^0 
	  \nonumber \\[1em]  
        & &  \qquad 
        + 5040 \rho^2 ( 1 + \rho^2 ) B_9^1
        + 105 \rho (24 -55 \rho^2 + 24 \rho^4) B_9^2 
	+ 2100 \rho^2 ( 1 + \rho^2) B_9^3 \nonumber \\[1em]
        & & \qquad \
	- 420 \rho ( 1 + \rho^2 + \rho^4 ) B_9^4 
        - 420 \rho^2 ( 1 + \rho^2) B_9^5  
	- 105 \rho^3 B_9^6 \Big] \,,
\end{eqnarray}

\begin{eqnarray}
c_{13}^{1} &=&  \frac{1}{16}  
  \Big[ -3 B_3^2 - \frac{135}{2} \rho^2 B_5^0 
    -  60 \rho B_5^1 - \frac{3}{2} ( 11 + 6 \rho^2 ) B_5^2 
    - 24 \rho B_5^3 + \frac{15}{4}  ( 60 + 109 \rho^2) B_7^0  \nonumber \\[1em]
    & &   \qquad 
    - \frac{15}{4} \rho ( -68 +21 \rho^2 )  B_7^1  
    - \frac{45}{2}  ( 2 + 11 \rho^2) B_7^2  
    - \frac{15}{4} \rho ( 20 + 13 \rho^2 ) B_7^3 
    - \frac{165}{4} \rho^2 B_7^4  \nonumber \\[1em]  
        & &  \qquad 
    - \frac{105}{8}  ( 12 + 59 \rho^2 + 33 \rho^4) B_9^0 
    + \frac{105}{8} \rho (34 + 43 \rho^2) B_9^1  
    + \frac{105}{2}  ( 1 + 4 \rho^2 + 7 \rho^4 ) B_9^2 
    \nonumber \\[1em]  
        & &  \qquad 
    - \frac{105}{16} \rho ( 4 + 19 \rho^2 ) B_9^3 
    - \frac{105}{8} \rho^2 ( 5 + 3 \rho^2) B_9^4 
    - \frac{315}{16} \rho^3 B_9^5 \Big] \, ,
\end{eqnarray}

\begin{eqnarray}
 c_{04}^{0} & = &  \frac{1}{192}  \Big[ -3 B_3^1  +  \frac{9}{2} \rho^{-1} ( \rho^2 + 30 ) B_5^0 
      + 180  B_5^1   + \frac{423}{2} \rho B_5^2  - 225 \rho^{-1} ( 5 \rho^2 + 2 ) B_7^0 
	\nonumber \\[1em] 
        && \qquad 
        + 45 \rho^2 ( -17 + 16 \rho^{-2}) B_7^1 + 1035 \rho  B_7^2 + 585 \rho^2 B_7^3  \nonumber \\[1em] 
        && \qquad 
        + \frac{315}{8} \rho^3 (59 + 56 \rho^{-2} +8 \rho^{-4}) B_9^0 - 315 \rho^4 ( 7 \rho^{-2} +4 \rho^{-4}) B_9^1 \nonumber \\[1em] 
        && \qquad - \frac{315}{2} \rho^3 (15 +2 \rho^{-2}) B_9^2
      + 945 \rho^2 B_9^3 + \frac{2835}{8} \rho^3 B_9^4 \Big] \,.
\end{eqnarray}

\noindent
The expansion in the text, equation~(\ref{phiexp}), is expressed in terms of the eccentricity vectors:

\beq 
{\bf e} \;=\; (e\cos g\,, \,e\sin g\,)\;;\qquad\quad {\bf e'} \;=\; (e'\cos g'\,, \,e'\sin g'\,)\,,
\eeq

\noindent
and neglects terms that are independent of ${\bf e}$. With these constraints, equation~(\ref{psi-4th}) reduces to:

\begin{eqnarray}
\Psi & = & \frac{\rho}{2{\rm max} (a,a')} \Big\{c_{20}^{0} e^2  +  c_{11}^{1} {\bf e}\cdot{\bf e'} + c_{40}^{0} e^4 +  c_{31}^{1} e^2 \left({\bf e}\cdot{\bf e'}\right) \nonumber \\[1em]
      & &   + (c_{22}^{0} - c_{22}^{2}) e^2 e'^2  + 2 c_{22}^{2} \left({\bf e}\cdot{\bf e'}\right)^{2} + c_{13}^{1} \left({\bf e}\cdot{\bf e'}\right) e'^2 \Big\}\,.
\end{eqnarray}

\noindent
By further specializing to rings of equal semi--major axis, i.e. $a = a'=a_0$, $\rho=1$ and $r = b/a_0$, and noting that in that case $c_{31}^{1} = c_{13}^{1}$, we can identify coefficients (Eqn.~\ref{phiexp}) with coefficients in the expansion above:

\beqa
\alpha &=& \frac{1}{2}\sqrt{\frac{G}{M_\bullet a_0^3}}\; c_{20}^{0}\,,\qquad 
\beta \;=\; \frac{1}{2}\sqrt{\frac{G}{M_\bullet a_0^3}}\; c_{11}^{1}\,,\qquad \gamma \;=\; \frac{1}{2}\sqrt{\frac{G}{M_\bullet a_0^3}}\;\left[c_{22}^{0} - c_{22}^{2}\right]\,,\nonumber\\[1em]
\lambda &=& \sqrt{\frac{G}{M_\bullet a_0^3}}\; c_{22}^{2}\,,\qquad
\kappa \;=\; \frac{1}{2}\sqrt{\frac{G}{M_\bullet a_0^3}}\; c_{31}^{1}\,,\qquad \chi \;=\; \frac{1}{2}\sqrt{\frac{G}{M_\bullet a_0^3}}\; c_{40}^{0}\,.
\eeqa

\noindent
We end with a brief remark about the convergence of the series expansion. The unsoftened expansions in powers of eccentricity are known to have a finite radius of convergence in the case of non--intersecting rings, in addition to blowing up when rings intersect. The later is alleviated by softening interactions as we have done. However, convergence of the softened series remains an issue: in the overlapping configurations considered here ($a = a' = a_0$), and for a given eccentricity $e$, the softening has to be larger than a critical value $b_c = a_0 e$ for the series to converge.  More details on convergence analysis and accuracy of the 4th order expansion can be found in \citet{mt11}.
 
\end{document}